**Mid-Infrared Spectroscopy of Planetary Analogs: A Database for Planetary Remote Sensing**

**Corresponding Author:** Andreas Morlok, Institut für Planetologie, Wilhelm-Klemm-Str. 10, 48149 Münster, Germany. Email: morlokan@uni-muenster.de, Tel. +49-251-83-39069

Stephan Klemme, Institut für Mineralogie, Corrensstraße 24, 48149 Münster, Germany. Email: stephan.klemme@uni-muenster.de

Iris Weber, Institut für Planetologie, Wilhelm-Klemm-Str. 10, 48149 Münster, Germany. Email: sonderm@uni-muenster.de

Aleksandra Stojic, Institut für Planetologie, Wilhelm-Klemm-Str. 10, 48149 Münster, Germany. Email: a.stojic@uni-muenster.de;

Martin Sohn, Hochschule Emden/Leer, Constantiaplatz 4, 26723 Emden, Germany, Email: martin.sohn@hs-emden-leer.de

Harald Hiesinger, Institut für Planetologie, Wilhelm-Klemm-Str. 10, 48149 Münster, Germany. Email: hiesinger@uni-muenster.de

Joern Helbert, Institute for Planetary Research, DLR, Rutherfordstrasse 2, 12489 Berlin, Germany, Email: joern.helbert@dlr.de

**Abstract**

The MERTIS (MErcury Radiometer and Thermal Infrared Spectrometer) instrument onboard the ESA/JAXA BepiColombo mission will provide mid-infrared data, which will be crucial to characterize the surface mineralogy of Mercury. In order to interpret the results, we are creating a database of mid-infrared spectra. As part of a study of synthetic glasses which are to serve as analog materials for the interpretation of remote sensing and modeling data, we present mid-infrared data for analog materials of Mercury regolith, surface and mantle compositions. In addition, we provide data for similar analogs of Earth, Moon, Venus, and Mars rocks for a coherent picture.

The analog samples have been first characterized by optical microscopy, Raman spectroscopy and EMPA. Powdered size fractions (0-25 µm, 25-63 µm, 63-125 µm, and 125 -250 µm) were studied in reflectance in the mid-infrared range from 2.5 to 18 µm (550 to 2000cm$^{-1}$), additional micro-FTIR analyses were also obtained.

Results for the size fractions of the surface and regolith analogs for Mercury show typical features for amorphous material with Christiansen Features (CF) at 8 – 8.1 µm, Reststrahlen Bands (RB) at 9.8-9.9 µm, and the Transparency Feature (TF) at 12 µm. The six bulk silicate Mercury analogs have varying CF positions from 8.1 – 9 µm, with RB crystalline features of various olivines dominating in most samples. Similarly, bulk silicate analogs of the other planetary bodies show glassy features for the surface analogs with CF from 7.9 µm (Earth Continental Crust) to 8.3 µm (Lunar Mare), strong RB from 9.5 µm (Earth Continental Crust ) to 10.6 µm (Lunar Mare and Highlands). TF are usually very weak for the glassy analogs.

Bulk silicate anlogs for the other planetary bodies are again dominated by crystalline olivine features. Trends between SCFM (SiO$_2$/(SiO$_2$+CaO+FeO+MgO)) index reflecting polymerization, CF and RB positions, and the SiO$_2$ contents in previous studies are basically confirmed, but there is indication that several samples (Moon Mare and Highlands) do not follow the trend observed for low SiO$_2$ samples and the RB position in earlier studies. Comparison with a ground based mid-IR spectrum of Mercury demonstrate general similarity in band positions with glassy Mercury surface and regolith material, and a chondrite based model of the planet. However, no mix so far can explain all spectral features.

1. Introduction

Mercury, the innermost planet of the Solar System, is characterized by a surface consisting of large impact basins, but also volcanic deposits (Hörz and Cintala, 1997; Denevi et al., 2013). The planet was so far only visited by the NASA Mariner 10 and MESSENGER space probes, with a third mission, ESA/JAXA BepiColombo being launched in October 2018. The MESSENGER x-ray a, gamma-ray and neutron spectrometers allowed to map the surface which resulted in geochemically distinct regions based on e.g. their Mg/Si, Al/Si, Ca/Si, Fe/Si and S/Si ratios. Fe contents are below 2 wt%, while S shows enrichments up 4 wt% (e.g. Peplowski et al., 2015; Weider et al., 2015; Vander Kaaden et al., 2016).

Crystallization experiments on analog magmas show olivine, plagioclase, pyroxene, silica and glass (Charlier et al.2013, Namur and Charlier, 2016). Additional ground based spectroscopy could only provide information about larger surface regions, indicating mainly feldspar, pyroxene, olivine (e.g.Sprague et al., 2007 and 2009). Closest terrestrial analog rocks for the hermean surface material are Fe-poor basalts, komatiites and boninites (Nittler et al., 2011; Stockstill-Cahill et al., 2012; Charlier et al., 2013; Vander Kaaden et al, 2017).

A wavelength range so far not covered by earlier missions to map the surface mineralogy in detail is the mid-infrared range. The feature-rich mid-infrared range enables determination of the mineralogical composition of planetary surfaces via remote sensing. The many diagnostic bands are an advantage compared to VIS/NIR spectroscopy.

The Mercury Radiometer and Thermal Infrared Spectrometer (MERTIS) onboard the ESA/JAXA BepiColombo mission to Mercury will allow mapping spectral features of the mercurian surface in the mid-infrared from 7-14 µm, with a spatial resolution of ~500 m (Benkhoff et al., 2010; Hiesinger et al., 2010). For the interpretation of the remote sensing data, laboratory spectra of analog materials are aquired at the IRIS (InfraRed and Raman for Interplanetary Spectroscopy) laboratory in Münster (Weber et al., 2016; Morlok et al., 2016a,b, 2017a, b).

Amorphous material and glasses are typical for surfaces of terrestrial planets and their moons, where they are formed in impacts and by volcanism (Hörz and Cintala, 1997; Denevi et al., 2013; Goudge et al., 2014). In this context, glasses are of particular interest as an analog material for remote sensing databases. The investigation of how the spectra of the rocks are changed by the impact process in the mid-infrared is important for the interpretation of infrared data from planetary bodies.

This paper is a follow-up to an earlier publication focussing on analog glasses with compositions of particular surface regions on Mercury, based on remote sensing chemical data and models from the MESSENGER mission (Morlok et al., 2017a). The mid-infrared spectra of these synthetic glasses show features typical for amorphous material: A single, strong Reststrahlen Band (RB) at 9.5 µm - 10.7 µm. In addition to the RBs, the Christiansen Features (CF) and Transparency Features (TF) shift depending on the $SiO_2$ and MgO contents (e.g., Hamilton, 2000; Morlok et al., 2017a).

Earlier studies of natural and synthetic glasses for planetary remote sensing covering the range from basaltic to intermediate compositions show similar results, with a dominating RB in the 9.2 – 10.7 µm range (Basilevsky et al., 2000; Morris et al., 2000; Moroz et al., 2009; DuFresne et al., 2009; Minitti et al., 2002; Minitti and Hamilton, 2010; Morlok et al., 2017a). Similar studies in the visible and near-infrared range were made for glasses with Martian and Mercurian compositions by Cannon et al. (2015 and 2017). Glasses with Martian surface composition have been studied in the form of quenched martian soil analog JSC Mars-1, which was analyzed in the mid-infrared by Basilevsky et al. (2000), Morris et al. (2000), and Moroz et al. (2009) and with other analogs studied by Minitti and Hamilton (2010) and Farrand et al. (2016). Terrestrial impact glasses were analyzed by Thomson and Schultz (2002), Gucsik et al. (2004), Faulques et al. (2001), Fröhlich et al. (2013), and Morlok et al. (2016a, b). Martin et al. (2017)

studied lunar meteorites with high glass contents; Pieters (1998) studied some glassy lunar material in the mid-infrared. Generally, these studies showed simple spectra typical for amorphous silicates, dominated by a strong, broad silicate feature around 10 µm (and lower for mafic compositions).

The aim of this study is to obtain a coherent set of reflectance mid-infrared spectra of synthetic analogs representative for larger components (bulk silicate, larger surface areas) of terrestrial planets (Venus, Earth, Moon, and Mars ) in the spectral range of interest for the MERTIS instrument, but also for planetary remote sensing in general. In the current study, we supplement spectra of new materials based on models of both the whole surface and regolith components of Mercury, and finally the bulk silicate part of the planet. Based on these end members, the percentage of the distinct size fractions of a given material could be calculated in the future observed spectra of Mercury. Additional micro-FTIR spectra help to identify glassy material in samples that have high contents of crystalline material despite quenching (Morlok et al., 2017a).

Previous studies of the surface regolith on Mercury indicated a high abundance of grains smaller than 25 µm in size (e.g.Sprague et al., 2007). Decreasing grain size causes also a decrease in the intensity of reflectance bands. Therefore, the spectral study of different grain size fractions is important. In order to account for large grain size variations and porosity of surface regoliths, four size fractions (0-25, 25-63, 63-125, 125-250 µm) were analyzed. The spectrum of the finest size fraction (particulary smaller than 25 µm) is characterized by the TF, which is usually not recognizable in spectra of larger grain size fractions. Transparency features ocurr when fine grained material becomes optically thin, while volume scattering dominates the scattering process. Since RBs have low intensities in the spectra of the smallest grain size fractions, the resulting feature of the volume scattering results in an often dominating reflectance feature (Salisbury and Eastes, 1985; Salisbury and Wald, 1992; Salisbury, 1993; Mustard and Hayes, 1997).

1.1. Sources for Starting Compositions of Analog Mixtures

For the chemical starting compositions (or analog mixtures) of the synthetic analogs used in this study we used geochemical models, remote sensing, averages of in-situ analyses and laboratory studies of samples. Starting compositions are presented in Tab.1.

The chemical data used for the analog mixture of bulk silicate Mercury were derived from various models (for an overview see Taylor and Scott, 2006). High refractory endmember models with high Al, Mg and Ca contents but no Fe and volastiles (abbreviated as Mercury BS (Ref) in this study) were modeled by Goettel (1988), who also provided an opposite volatile-rich endmember with high Fe and alkalis (abbreviated as Mercury BS (Vol)). The model of Morgan and Anders (1980) used seven chondritic components and geophysical data (Mercury BS (Chon1)). Krot et al. (2001) used average bulk compositions of chondrules from metal-rich choindrites. The high MgO/low $SiO_2$ composition of Fegley and Cameron (1987) (Mercury BS (Vap)) is based on a vaporization model of a molten mantle of magma with chondritic composition. The ‚Mixture' model (Mercury BS (Mix)) averages geochemical models from several studies (Jagoutz et al., 1979; Taylor and McLennan 1985, Goettel, 1988; McDonough and Sun, 1995). Average global surface and regolith compositions (Mercury Surface and Mercury Regolith)(vander Kaaden et al., 2017) are based on the MESSENGER derived surface elemental chemistry of Mercury.

The analog mixture for the surface of Venus (Venus Surface) is the average of the chemistry determined from the Venera 13, Venera 14, and Vega 2 landers (Fegley et al., 2003). Since the bulk silicate Venus is probably very similar to the bulk silicate Earth (Earth BS) (Morgan and Anders, 1980), we use the latter as an analog. Analog mixtures for the chemistry of lunar mare and highland regions are based on averages of the Apollo and Luna samples (Warren, 2006). The bulk silicate Moon analog mixture (Moon BS) is from a model in O'Neill (1991) based on a giant impact formation of the moon, that uses a chemical mixture dominated by Earth's mantle and chondritic material.

The surface of the Earth is represented by an analog mixture based on a geochemical average of the various parts of continental crust (Earth Continental Crust) (Rudnick and Gao, 2006); and an average for laboratory analyses of Mid-Ocean Ridge Basalts (Earth MORB) (Klein, 2006). The analog mixture for Bulk silicate Earth (Earth BS) is based on mainly chondritic material (Palme and O'Neill, 2003).

Analog mixtures for Mars surface and regolith (Mars Surface and Mars Regolith) are from Santos et al. (2015), which are based on the chemical analyses of Martian meteorites. The data for the analog mixture of bulk silicate Mars (Mars BS) was modeled by Taylor (2013), using a combination of analyses of martian meteorites and geophysical models.

## 2. Samples and Techniques

### 2.1 Sample Synthesis and Preparation

In order to use the same compositions as in previous papers (which varies among the different sources for this study) we chose to limit the composition used in this study to $SiO_2$, $TiO_2$, $Al_2O_3$, FeO, MgO, CaO, $Na_2O$, and $K_2O$. Some chemical components with an average content below 0.5 wt.% (e.g., Cr2O3) were omitted for simplification. Oxide and carbonate mixtures were ground to a fine powder in an agate mortar under acetone and then dried. The resulting mixtures were slowly heated to 1000°C to de-carbonate in medium sized Pt crucibles. Following this, the mixtures were melted in a conventional box furnace at 1450°C for 2h. The crucibles were swiftly taken out of the furnace and submerged in water. The samples were vitrified within 10 secs. Oxygen fugacity was not controlled, since the samples were melted in a box furnace in air.

The resulting run product was crushed in a steel mortar, the powder cleaned in acetone and dry sieved into four size fractions: 0-25 μm, 25-63 μm, 63-125 μm, and 125-250 μm, by using an automatic Retsch Tap Sieve. In order to remove clinging fines, the larger two fractions were again cleaned in acetone. In addition, polished thick sections from the pure glass sample were prepared for microscopic investigation. Crushing and sieving of bulk rocks consisting of minerals with varying physical properties (e.g. hardness) may affect the modal composition of the size fractions. This has to be kept in mind when discussing the spectral results of the size fractions.

## 2.2. Optical Microscopy

Overview images to identify the crystalline or amorphous character of the samples were made using optical microscopy (Fig. 1). Images of all polished sections were made with a KEYENCE Digital Microscope VHX-500F.

## 2.3. Raman

All Raman measurements were conducted using an Ocean Optics IDR-Micro Raman system at the IRIS laboratory in Münster, operating with a OneFocus optical system equipped with a 40 x objective. The laser excitation is 532 nm and the resolution is about 7 cm$^{-1}$ (=distance between 2 detector pixels). Spectra were obtained with a laser power of 1.8 mW, which corresponds to a laser irradiation of 0.6 mW/µm$^2$ on the sample. The spot size on the sample is approximately 2 µm in diameter. Every spectrum is the result of 4 measurements at 3 seconds acquisition time.

All spectra are automatically background- and baseline-subtracted and have not been smoothed. The height of a Raman signal corresponds to the quality of the individual Raman scatterer itself (e.g., the double peak in olivine appears because the two SiO$_4$-stretching ($v_1$ and $v_3$) modes are active), therefore all spectra are given with arbitrary units (Fig. 2).

## 2.3. Electron Microprobe Analysis (EMPA)

Detailed quantitative analyses of the glasses and the phases crystallized during the quenching process were made with a JEOL JXA-8530F Hyperprobe electron probe micro analyzer (EPMA) equipped with five wavelength dispersive spectrometers (WDS) (Tab. 1). For the analyses, the electron micro probe was operated at an excitation voltage of 15 kV and a beam current of 5 nA. The beam diameter was defocused to 5 µm. The counting time was 5 seconds on the peak and 2 seconds on the background of each element, respectively. For the listed mineral analyses, we used an excitation voltage of 15 kV and a beam current of 15 nA with a slightly defocused beam diameter of 2 µm. The counting time for Mg, Al, Si, Ca, Fe, Ti, Cr, and Mn was 15 seconds on the peak and 5 seconds on the background. In order to avoid loss of the volatile elements Na and K, the counting time for the elements was reduced to 5 seconds on the peak and 2 seconds on the backgroun.

The following natural and synthetic minerals with well-known compositions were used as standards: Jadeite ($Na_2O$), SanCarlos olivine (MgO), disthene ($Al_2O_3$), hypersthene ($SiO_2$), sanidine ($K_2O$), diopside (CaO), fayalite (FeO), rutile (TiO2), $Cr_2O_3$ and rhodonite (MnO). In addition, backscattered electron (BSE) images (Fig. 3) show crystalline phases (brighter skelletical phases) formed during quenching.

## 2.5. Bulk FTIR

The powdered samples were filled into aluminum sample cups, one cm diameter. The surface was gently flattened with a spatula following a procedure analogue to that described by Mustard and Hayes (1997). For the bi-directional analyses of the sieved bulk powder size fractions in the wavelength region of 2.5-18 µm, we used a Bruker Vertex 70 infrared system with a cooled MCT detector at IRIS laboratory. For a high signal- to noise-ratio, analyses were made under low pressure ($10^{-3}$ bar). Each spectrum consists of 512 accumulated scans for each size fraction, with a spectral resolution of 0.02 µm. The machine background was removed using a diffuse gold standard (INFRAGOLD[TM]). The wavenumber accuracy is 0.01 cm$^{-1}$. In order to emulate various observational geometries of an orbiter, analyses were made with a variable geometry stage (Bruker A513). The data used in this study were made at 20° incidence (i) and 30° emergence angle (e).

The results are presentes in reflectance from 6-18 µm (Fig. 4 and 5). The spectral range of the MERTIS spectrometer is from 7-14 µm, but features at shorter and longer wavelengths can be of interest for other studies. Band positions for both powder and microscope analyses were obtained using Origin Pro 8. The wavelength of a specific feature was determined by the position of thestrongest reflectance, or, for the CF, of the lowest reflectance.

In the interpretation of remote sensing data from MERTIS, laboratory reflectance and emission data will to be compared. This is done by Kirchhoff's law: $\varepsilon = 1 - R$ (R = Reflectance, ε = Emission) (Nicodemus, 1965). However, a requirement for using Kirchhoff's law is that the reflected light in all directions has to be collected. This works best when comparing directional emissivity and directional hemispherical reflectance. In our study, a bi-directional variable mirror set-up (Bruker A513) was used – we did not messure directional hemispherical reflectance. This has to be kept in mind when comparing the results in a quantitative manner with emission data (Salisbury et al., 1991; Hapke, 1993; Thomson and Salisbury, 1993; Salisbury et al., 1994; Christensen et al., 2001). The spectra presented in this study are soon accessible via an online database at the Institut für Planetologie in Münster (http://www.uni-muenster.de/Planetology/en/ifp/ausstattung/iris_spectra_database.html), and the Berlin Emissivity Database (BED).

**2.6. Micro FTIR**

A Bruker Hyperion 2000 IR microscope attached to the external port of a Bruker Vertex 70v at the Hochschule Emden/Leer was used for the analyses of selected areas on polished sections. A 256×256 µm$^2$ sized aperture was used to analyze specific areas and features with in situ reflectance spectroscopy.

128 scans were integrated for each spectrum, and a gold mirror was used for background calibration (Fig. 6).

## 3. Results

### 3.1. Sample Characterization (Optical, EMPA, Raman)

Results of the chemical composition of the synthetic glass are presented in Tab. 1a and b. The six analog glasses based on bulk silicate Mercury models all show probably signs of crystallization of small phases in the optical images (Fig. 1a and b). EMPA analyses (Tab. 1a) show that mainly forsteritic olivine has crystallized for Mercury BS (Vol), Mercur BS (Mix), Mercury BS (Chon1), Mercury BS (Chon2), and Mercury BS (Ref). Mercury BS (Vap) is dominated by Ca rich monticellite, which is also confirmed by Raman spectroscopy (Fig. 2a) (McMillan and Piriou 1982; Chopelas et al., 1991; DiGenova et al., 2015). The Mercury BS (Ref), Mercury BS (Vol) and Mercury BS (Chon1) samples also show spinel, magnetite, and magnesioferrite, which was also detected by Raman spectroscopy (Fig.2b; Tab.1b). Mercury BS (Vap) also contains some periclase (MgO), probabaly unreacted starting material.

Potential glass phases with strong chemical fractionation due to crystal formation compared to the bulk starting material were probably identified in all the Mercury model samples based on their location between crystals, but it was difficult to obtain useful analyses due to the small size and intermixed nature of the spots. Exceptions were Mercury BS (Vap) and Mercury BS (Vol).
The two synthetic analog samples for the mercurian surface, Mercury Surface and Mercury Regolith, on the other hand, include homogenous glass, with very few crystallites (Fig.1b). Chemical compositions close to the starting mixture indicate glass, since crystallizing phases would have changed the bulk composition of the remaining glass (Tab.1a).

The Venus Surface analogue sample shows some inclusions in homogeneous material with a chemical composition relatively close to the starting material (Fig. 1a, Tab. 1a). Both the Moon mare and highland samples are quite homogeneous in the optical images (Fig.1b), with chemical compositions of the main elements very close to the starting material, confirming a mainlyglassy phase (with the exception of $TiO_2$ which is fractionated in the highland sample) (Tab. 1a). The BS Moon analog also shows abundant crystals in the SEM/BSE image (Fig.1b), dominated by forsteritic olivine but also with some pigeonite or fassaite content and minor spinel phases (spinel, magnesioferrite, magnetite) (Fig. 1; representative data in Tab. 1a-c) (Deer et al., 1992).

The analogue for Earth BS (basically Earth primitive mantle) also exhibits many crystals in a glassy matrix in the SEM/BSE images (Fig.1b). EMPA analyses show manly forsteritic olivine, with minor spinel phases (spinel, magnesioferrite, magnetite) (Tab. 1a-c). The analog for the oceanic basalts (Earth MORB) is very glassy in the optical image, and has a chemistry very close to the starting material. Similar, the synthetic Earth Continental Crust sample consists of homogeneous glass with a chemistry of the main elements close to the starting mixture (Fig. 1; Tab. 1a-c).

The Mars BS sample shows crystalline fatures in addition to quench textures in the SEM/BSE images (Fig.1b), EMPA analyses show mainly forsteritc olivine, and fine dendritic pigeonite or fassaite (Fig. 3). Also, minor spinel phases were found in this sample (Deer et al., 1992). The Mars Regolith and Mars Surface analogs are also glassy in the optical images, and have chemical compositions similar to the starting mixture (Fig. 1; Tab. 1a).

### 3.2. FTIR Spectroscopy

#### 3.2.1 Bulk Powders

Band positions for the analyses of the powdered samples are presented in Tab. 2a-c, and the spectra are shown in Fig. 4 and 5. The material based on the model compositions of the bulk silicate Mercury show feature-rich spectra (Fig. 4a). The increased number of RBs, as well as the BSE images (Fig.1b) indicates that crystals formed in the melt during the quenching. At shorter wavelengths, with exception of Mercury BS (Vol), they all have volatile features at 2.8 µm – 3 µm (water), 3.4 µm and 3.5 µm (hydrocarbons) (Salisbury et al., 1988). Further features, possibly overtones of silicates (Bowey and Hofmeister, 2005), are found between 4 and 6.5 µm (Tab. 2a).

Three samples – Mercury BS (Vol), Mercury BS (Mix), Mercury BS (Chon1) (Fig. 4a) - show similar spectra with the CF located between 8.1 µm and 8.8 µm. The Mercury BS (Mix) sample exhibits the strongest shift of the CF (8.7 µm- 8.8 µm). Significant RB occur at 9.8 – 9.9 µm, 10.1 µm -10.2 µm, 10.6 µm – 10.7 µm, 11.1 µm -11.2 µm and 11.9 µm, plus a band at longer wavelengths at 16.1 µm -16.3 µm. The TF is between 12.4 µm and 12.5 µm (Tab. 2a). The Mercury BS (Chon2) (Fig. 4a) sample has a similar shape with a CF positioned at 8.2 µm -8.4 µm, but also strong RBs at 9.3 µm and 10.5 µm, plus weak bands at 10.14 µm, 10.8 µm, 13.6 µm, and 14 µm. The TF is at 12.4 µm (Tab. 2a). These crystallline features can be explained mostly with olivine bands (Hamilton et al., 2010).

The Mercury BS (Vap) (Fig. 4a) spectrum shows a different band shape compared to the other Mercury BS spectra, with strong RBs at 9.5 µm and 10.7 µm, and minor features at 10.3 µm, 13.9 µm, 14.3 µm-14.4 µm, and 16.3 µm – 16.4 µm. The CF is at 8.9 µm - 9 µm and the TF at 12.6 µm (Tab. 2a). These features indicate a mix of olivine and monticellite bands (Hofmeister, 1997; Hamilton, 2010).

The Mercury BS (Ref) (Fig. 4a) shows a significantly diverging spectrum, with a broad spinel feature (Chihara et al., 2000; Fabian et al., 2001) from 12.9 µm to a broad peak at 13.7 µm - 14.2 µm. There are also strong olivine RBs at 10.5 µm – 10.6, 9.3 µm, 10.1 µm, 11.4 µm, and 11.9 µm (Hamilton, 2010). The CF is located at 8.6 µm, and the TF is difficult to locate due to the strong RB band in the region (Tab. 2a). The Mercury Surface and Mercury Regolith analog samples (Fig. 4b) show typical features of glassy material, i.e. a single, dominating RB at 9.8 µm – 9.9 µm, a CF at 8 µm – 8.1 µm, and the TF at 12 µm. There are also less prominent features at shorter wavelengths, i.e. volatile bands at 2.8 µm -2.9 µm, 3.4 µm and 3.5 µm (Tab.2a).

The sample Venus Surface (Tab. 2b; Fig. 4b) also has amorphous features with a RB at 10.5 µm – 10.6 µm. The CF is at 8.1 µm, the TF is too weak to pinpoint the position. Again, features below 8 µm are rare and only represent the volatile bands. Similar to the Mercury bulk silicate models, the Moon BS (Tab. 2b; Fig. 4b) shows crystalline features with RBs at 10.1 µm, 10.8 µm -11.3 µm, 11.9 µm, and 16.2 µm -16.3 µm. The CF is found at 8.2 µm - 8.5 µm, the TF at 12.3 µm. At shorter wavelengths, there are only very weak volatile features, but bands occur 4.9 µm, 5.5 µm, 5.6 µm, and 6 µm. The strong crystalline bands correspond to olivine features (Hamilton et al., 2010). The spectra of samples with Moon highland and mare compositions (Fig. 4b) show again glassy features with the RB at 10.6 µm for the mare and at 10.5 µm -10.6 µm for the highland analogs. The CF are at 8.2 µm – 8.3 µm (mare) and 8.2 µm (highlands). The latter sample also has features at 14.6 µm – 14.9 µm. Both spectra have no clearly identifiable TF (Tab.2b).

Earth BS (Tab.2c; Fig.4c) also shows mostly olivine crystalline features with RB at 10.1 µm and 10.6 µm – 11.2 µm, plus weaker bands at 9.7 µm (shoulder), 11.9 µm and at 16.2 µm – 16.4 µm (Hamilton, 2010). The TF is located at 12.5 µm, the CF between 8.2 µm and 8.6 µm. In addition to the volatile bands there are features at 4.9 µm, 5.6 µm and 6 µm.

Both Earth MORB and Earth Continental Crust (Tab. 2c; Fig. 4c) samples have again glassy features. The main RB is at 10.2 µm for the Earth MORB glass, and 9.5 µm for the Earth Continental Crust analog. TFs occur at 11.8 µm (Earth MORB) and 11.7 µm (Earth Continental Crust). The CF is found between 8.1 µm and 8.2 µm for Earth MORB and at 7.9 µm for the Earth Cont.Crust analog. At shorter wavelengths, both samples show the usual volatile bands between 2.8 µm and 3.5 µm.

Similar to the other samples, the Bulk silicate Mars (Mars BS) has mainly crystalline olivine features (Tab. 2c; Fig. 4c)(Hamilton, 2010). Strong RBs are visible at 10.1 µm, and 10.7 µm -11.3 µm. Weaker features occur at 9.7 µm - 9.8 µm (shoulder), 12 µm and 16.4 µm. The TF is at 12.4 µm, and the CF is at 8.0 µm - 8.3 µm. Volatile bands are weak, at shorter wavelengths there are features at 4.9 µm, 5.6 µm, and 6 µm. Analog spectra Mars Surface and Mars Regolith again show typical amorphous features with a dominating RB band at 10 µm (Mars Surface) and 9.7 µm (Mars Regolith). The CF and TF are at 8.2 μm / 11.9 μm (Mars Surface) and 8 μm / 11.9 μm (Mars Regolith), respectively. At shorter wavelegths, there are volatile bands from 2.8 μm to 3.5 μm (Tab.2c).

### 3.2.2. Micro FTIR

Spectra for the Moon highlands (Fig. 6; Tab. 3) showed very similar results for the analyzed spots: a CF at 8.2 μm and a dominating RB at 10.6 μm, very similar to the powder study. The situation is similar for the Moon mare in-situ analyses: the two representative spots (Fig. 6; Tab. 3) have a CF at 8.3 μm and the main RB at 10.6 μm. Similarily, the spots for the Venus Surface material does not vary much between the spots, while the CF is at 8.1 μm, the strong RB is at 10.6 μm. Again, all the Micro-FTIR analyses are very similar to the bulk powder analyses and indicate a very high homogeneity of the glass. In contrast to the bulk powder measurements, no significant features at wavelengths shorter than the CF were visible.

## 4. Discussion

### 4.1. Comparison between Chemical and Spectral Features

Of particular interest for the comparison of the results of our study with remote sensing data of Mercury and other bodies is the identification of characteristic spectral features. One spectral feature that plays a central role is the CF, since it is relatively easy to identify as the reflectance minimum (or emission maximum) even in data with low spectral contrast. Even more, the correlation with the bulk chemical composition of the sample (e.g., as proxy for the $SiO_2$ content) makes this feature particularly interesting (e.g., Salisbury, 1993). In Fig. 7 we compare the CF positon of the glassy samples with the $SiO_2$ contents. The results of this study fall close to the trendline for powdered crystalline terrestrial rocks (Cooper et al., 2002). They are also very similar to the spectral signatures of synthetic glasses with the compositions of surface regions of Mercury (Morlok et al., 2017a), but are mostly different from the features of impact glasses from the terrestrial surface, which tend to have more SiO$_2$-rich compositions (Morlok et al., 2016b). The results for Moon mare and highland regions extend the range of data into the low SiO$_2$/high CF region.

The single, strong RB of the glassy samples could also be a spectral feature easy to identify (Fig.8) in remote sensing data. Again, the results of this study mostly fall in the region of the earlier studies of glasses with Mercury surface composition (Morlok et al, 2017a), and synthetic glasses with basaltic composition (DuFresne et al., 2009). Some samples at the low SiO$_2$/high CF basaltic end of the range follow a trend already observed by DuFresne et al. (2009) and Morlok et al. (2017a). This trend contains a horizonal 'bend' near 50 wt% of SiO$_2$. Only the Moon Mare and Moon Highland samples (both bulk and micro-FTIR) do not follow the 'horizontal' trend. They extend linear trend seen for most of the spectra of materials with higher SiO$_2$ contents.

A way to describe the degree of polymerization of the involved material (based on the interconnection of the SiO$_4$ tetrahedra) is the SCFM index (Fig. 9). This index is calculated as SiO$_2$/(SiO$_2$+CaO+FeO+MgO) (Walter and Salisbury, 1989), and has been used for spectral studies of e.g. lunar samples (e.g. Salisbury et al., 1997; Isaacson et al., 2010) or martian analogs (e.g.Wyatt et al., 2001; Cloutis et al., 2002). Again, the results of this study are similar to the results for Mercurian glasses (Morlok et al., 2017a) and terrestrial impact glasses (Morlok et al., 2016b). Most of the results plot off the trendline for crystalline terrestrial rocks (Cooper et al., 2002). This divergence from the trendline for crystalline materials could make this method useful to identify glasses in remote sensing data. Results for crystalline lunar material in Isaacson et al.(2010) and lunar Apollo soils (Salisbury et al., 1997) followed the trend observed in Cooper et al., 2002.

In general, crystalline features appear in samples with an initial MgO content of over 30 wt% in the starting material (Tab.1a). All spectra of synthetic analog glass with lower initial MgO contents (in this study ranging from 4.7 to 18.3 wt%; Tab.1a) show barelay any sign of crystallization. This critical

MgO content confirms the findings of Morlok et al. (2017a) of synthetic glass with the composition of specific surface regions of Mercury. Here, a treshold for crystallization was suggested at about 23 wt% MgO. The impact of high MgO (and FeO) contents on olivine and pyroxene quench crystals was observed in earlier studies on comparable mafic mixtures (e.g.Donaldson, 1979; Herzberg and Zhang, 1998). Dendritic crystallization patterns as observed in the high MgO samples (Fig.1b) are also commonly found in materials formed under non-equilibrium conditions like quenching (e.g. Jambon et al., 1992; Zhao et al., 2011).

**4.2. Mid-Infrared Spectra of Mercury**

The few available ground-based spectra of Mercury in the the mid-infrared cover larger surface regions ($10^4$-$10^6$km$^2$). These spectroscopic observations indicate a composition dominated by plagioclase and some pyroxene (Donaldson-Hanna et al., 2007; Sprague et al., 1994, 2000, 2002, 2007; Sprague and Roush, 1998; Emery et al., 1998; Cooper et al., 2001). Owing to observational problems, most terrestrial spectra of Mercury offer only weak spectral contrast and low signal to noise ratio e.g. (Sprague et al., 2007). Therefore, we limit the comparison of the spectra of our synthetic analogs with only one high signal to noise spectrum of Mercury. This spectrum was obtained by the Mid-Infrared Camera (MIRAC) at the Kitty Peak Observatory of a region at about 210-250° longitude (Sprague et al., 2000). The chemical and mineralogical variation in surface composition based on the MESSENGER data (e.g. vander Kaaden et al., 2016) shows strong variation. So the spectra will probably only represent a part of the surface mineralogy of Mercury. The baseline-subtracted spectrum (recalculated from emission) has strong RBs at 9.3 µm, 9.9 µm, and 11 µm (Fig. 10). The CF is at 8.5 µm, and a potential TF occurs at 12.4 µm.

A comparison of the Mercury spectrum with synthetic analogs based on the chemistry for distinct surface regions of Mercury (Morlok et al., 2017a) showed that not all spectral features could be reproduced. The best matching mixture obtained was of a 125-250 µm fraction of the High- Aluminium

Region (vander Kaaden et al. 2016) and the High Magnesium Region (vander Kaaden et al. 2016). Here, no equivalent for the 9.3 µm feature in the Mercurian spectrum was found, and band shapes differed.

If we add the Mercury analog spectra of our current study, the 0-25 µm Mercury BS (Chond2) spectrum would reproduce the 9.3 µm feature and the TF very well. Also, the CF would be close to that seen in the telescopic data (8.2 µm-8.4 µm). Unfortunately, there are no equivalents for the 9.9 µm and 11 µm bands in the Mercury spectrum (Sprague et al., 2000). A representative of the spectra with high crystalline olivine content, the 0-25 µm Mercury BS (Vol) spectrum would have a broad band in the region of the 11 µm band in the Mercurian spectrum (but with a much broader band shape), and a TF at 12.4 µm. With the glassy 125 – 250 µm Mercury Surface and Mercury Regolith spectra, the strong RB at 9.8 µm – 9.9 µm would reproduce the 9.9 µm band of the remote sensing Mercury spectrum, but the shape of the bands in the analog spectra are much broader.

Obviously, the surface spectrum cannot be explained with glassy material alone. Particularily, the areas formed in volcanic events will probably also contain abundant crystalline phases. So while glasses can be expected to form a important component of the surface regolith, with up to 45% agglutinates (Warell et al., 2010), further crystalline components such as volcanic rocks have to be taken into account in mixing calculations. Also, this comparison is based merely on peak matching, for detailed an quantitative studies the full pattern and band shape has to be taken into account.

## 5. Summary and Conclusions

In this study we presented the mid-infrared spectra of synthetic analogs for surface regions, regoliths and bulk silicate parts of terrestrial bodies in the inner Solar System based on remote-sensing data and modeling. The resulting glassy or amorphous samples with MgO contents of up to 18.3 wt.% show a dominating RB at 9.5 µm (Earth Continental Crust) to 10.6 µm (Moon Mare and Highlands, Venus Surface). This is basically the same range observed for analog glasses of surface regions for Mercury

(Morlok et al., 2017a), and also overlaps with other earlier studies of glassy planetary analog materials with RB in the 8.9 – 10.5 µm range (Faulques et al., 2001; Thomson and Schultz, 2002; DuFresne et al., 2009, Minitti et al., 2002 and Minitti and Hamilton, 2010; Gucsik et al., 2004; Fröhlich et al., 2013 and Morlok et al., 2016b). Micro-FTIR studies of Moon analog samples essentially confirm theses findings and demonstrate the homogeneity of the glass.

Another group, representing the highly mafic bulk silicate components of the bodies, shows strong signs of crystallization. Spectral features typical for forsteritic olivine were observed for samples with higher MgO contents. Additional mineral species were spinel, dominating a highly refractory model, Mercury BS (Ref) (Goettel et al., 1988), montecellite, pigeonite or fassaite and further minor spinel phases like magnesioferrite and magnetite-rich spinel solid solutions.

When put into context, the spectral features of our newly measured planetary glasses behave similar to the results of earlier studies. The glassy materials reflects the bulk composition when the positions of the CF is put into relation to the $SiO_2$ content. Similarly, the positions of the main RB of the amorphous material shows a correlation with the $SiO_2$ content. However, there is indication that some very mafic samples (Moon mare and highlands) do not follow the trend observed for low $SiO_2$ samples in our earlier and other studies (DuFresne et al., 2009; Morlok et al., 2017a).

Finally, the results of this study confirm a trend observed in the comparison of the polymerization of the sample in form of the SCFM index with the CF. Again, the results for the glassy material plot off the trendline for crystalline terrestrial material. Thus, this behavior might be useful to identify glassy material in remote sensing data. However, basic terrestrial rocks in earlier studies show a similar behavior (Cooper et., 2002), so this might limit the use of this comparison to clearly identify amorphous material.

Finally, we compared the spectra of our synthetic planetary analogs with a ground-based mid-IR spectrum of Mercury and found similarity in band positions to glassy Mercury surface and regolith materials, and a chondrite-based BD model of the planet. However, no mix of analog material is a perfect match for all spectral features observed on the surface of Mercury.

**Acknowledgements**

Many thanks to Ulla Heitmann (Münster) for sample preparations and Isabelle Dittmar (Emden) for analytical support. This work was partly supportet by DLR grant 50 QW 1302/1701 in the framework of the BepiColombo mission.

**Figure Captions**

Figure 1a: Optical reflected light images of polished blocks of the analog samples used in this study. Squares: 256×256 µm sized areas analyzed with micro-FTIR.

Figure 1b: Optical reflected light images (top ) and SEM/BSE images (bottom) of enlarged parts of several samples. Mercury Regolith and Surface, Moon Mare and Highlands: Optical reflected light images. No crystals are visible, all features are scratches. Mercury, Mars, Earth and Moon BS: SEM/BSE images. Here several tens of micron sized crystals are visible are visible in the brighter glassy groundmass. Earth and Mars BS also show fine, bright sceletal quenching structures.

Fig. 2: Raman spectra for crystalline inclusions in the glasses Mercury BS (Vap) and (Ref). (a) Monticellite in Mercury BS (Vap), (b) Magnesioferrite and olivine in Mercury BS (Ref). (Compare: McMillan and Piriou 1982; Chopelas et al., 1991; DiGenova et al., 2015).

**Figure 3:** EMPA backscatter electron image of area in the Mars BS sample that contains typical phases observed in crystallized samples. a) olivine b) mix of dendritic pigeonite/fassaite with olivine c) magnesioferrite and magnetite.

**Figure 4a-c:** FTIR reflectance spectra of powdered glasses in four size fractions (0-25 µm, 25-63 µm, 63-125 µm, 125-250 µm). (a) Analog samples with compositions based on models of bulk silicate Mercury, (b) samples with Mercury surface and regolith composition, Venus Surface and Moon BS, (c) Analog samples with Earth BS, Earth Continental Crust and ocean basalt (Earth MORB) composition; and Mars BS, Mars Regolith and Mars Surface composition.

**Figure 5a:** FTIR reflectance powder spectra of the 125-250 µm size fraction also presented in Fig.4a-c shown together for comparison (in µm).

**Figure 5b:** Representative FTIR reflectance spectra at wavelengths shorter 7 µm. Typical features for water between 2 and 4 µm and overtone bands between 4.5 and 6.5 µm are marked with vertical lines. Continous line: bulk silicate analogs, dotted line: spectra of surface materials.

**Figure 6:** Micro-FTIR reflectance spectra of polished samples for Venus Surface and Moon Mare and Moon Highlands. Analyzed spots are marked in the optical images of the samples (Fig.1a).

**Figure 7:** $SiO_2$ contents of glasses (in wt.%) vs. position of the Christiansen Feature (in µm). The band positions of the analog samples of this study fall on the dashed trend line defined by earlier studies on impact glasses (Morlok et al., 2016b and 2017a) and crystalline terrestrial rocks (Cooper et al., 2002).

**Figure 8:** $SiO_2$ contents of glasses (in wt.%) vs. the position of the strongest Reststrahlen-Band (in µm). Band positions of analog glasses from this study are similar to those of earlier studies, showing a linear trend (DuFresne et al., 2009; Lee et al., 2010; Morlok et al., 2016b, 2017a), but also diverge from the trend at lower $SiO_2$ content.

**Figure 9:** Comparison of the SCFM index ($SiO_2/(SiO_2+CaO+FeO+MgO)$) (Walter and Salisbury, 1989), with the position of the CF (in µm). Band positions for analog glasses in this study confirm the trends in Morlok et al. (2016b; 2017a): i.e. synthetic glasses plots below the trend line for crystalline phases. Dotted trend line is for terrestrial rocks (Cooper et al., 2002).

**Figure 10:** Comparison of mid-infrared spectra of synthetic analog glasses with compositions for specific regions of Mercury with a mid-infrared spectrum of the planet from ground based observation (Sprague et al., 2000). The spectrum of the finest size fraction of the high Mg-region (HMC VdK) (Morlok et al., 2017a) provides good fits for the Transparency Feature (TF). The finest fraction of a chondrite-based Bulk Silicate Mercury spectrum, BS Mercury (Chond2) has good fits for the CF, TF, and the feature at 9.3 µm. Mercury Surface and Regolith spectra have their strong Restrahlen Band (RB) close to the 9.9 µm

feature, but band shapes are different. The band at 11 μm has no good equivalent among the synthetic analogs so far.

|  | Merc. BS (Mix) |  | Start | Merc. BS (Chon1) |  | Start | Merc. BS (Chon2) |  | Start | Merc. BS (Ref) |  | Start |
|---|---|---|---|---|---|---|---|---|---|---|---|---|
| Na$_2$O | 0.09 | 0.02 | 0.00 | 0.03 | 0.02 | 0.00 | 0.08 | 0.03 | 0.00 | 0.68 | 0.42 | 0.00 |
| MgO | 13.94 | 1.02 | 36.10 | 17.16 | 0.47 | 33.70 | 16.65 | 0.18 | 37.40 | 9.84 | 2.32 | 34.58 |
| SiO$_2$ | 38.80 | 0.22 | 37.60 | 52.99 | 0.34 | 47.10 | 55.98 | 0.18 | 50.80 | 65.12 | 8.17 | 32.58 |
| Al$_2$O$_3$ | 16.40 | 0.45 | 11.70 | 11.86 | 0.19 | 6.40 | 12.34 | 0.08 | 5.20 | 17.88 | 2.66 | 16.62 |
| K$_2$O | 0.06 | 0.02 | 0.00 | 0.04 | 0.01 | 0.00 | 0.11 | 0.00 | 0.00 | 0.05 | 0.07 | 0.00 |
| CaO | 24.27 | 0.63 | 10.70 | 10.24 | 0.08 | 5.20 | 11.44 | 0.20 | 3.60 | 0.46 | 0.54 | 15.19 |
| FeO | 4.78 | 0.04 | 3.00 | 6.31 | 0.08 | 3.70 | 2.60 | 0.06 | 2.00 | 0.00 | 0.00 | 0.00 |
| TiO$_2$ | 1.14 | 0.07 | 0.50 | 0.54 | 0.09 | 0.30 | 0.43 | 0.10 | 0.20 | 0.00 | 0.00 | 0.72 |
| Total | 99.48 |  | 99.60 | 99.17 |  | 96.40 | 99.63 |  | 99.20 | 94.03 |  | 99.69 |
| N | 3 |  |  | 3 |  |  | 3 |  |  | 3 |  |  |

|  | Merc. (Rego) |  | Start | Merc. (Surf) |  | Start | Venus Surf. |  | Start | Moon High |  | Start |
|---|---|---|---|---|---|---|---|---|---|---|---|---|
| Na$_2$O | 3.81 | 0.11 | 3.58 | 3.44 | 0.07 | 3.51 | 0.06 | 0.03 | 0.00 | 0.06 | 0.03 | 0.00 |
| MgO | 17.15 | 0.18 | 18.28 | 17.51 | 0.34 | 17.94 | 11.06 | 0.07 | 10.33 | 5.77 | 0.12 | 5.70 |
| SiO$_2$ | 58.21 | 0.13 | 56.94 | 56.09 | 0.39 | 55.89 | 50.07 | 0.20 | 46.47 | 44.76 | 0.21 | 44.90 |
| Al$_2$O$_3$ | 14.83 | 0.11 | 14.32 | 14.55 | 0.15 | 14.06 | 17.93 | 0.04 | 16.57 | 26.89 | 0.26 | 26.90 |
| K$_2$O | 0.05 | 0.02 | 0.00 | 0.05 | 0.02 | 0.00 | 0.05 | 0.01 | 0.00 | 0.04 | 0.02 | 0.00 |
| CaO | 6.03 | 0.05 | 5.85 | 6.04 | 0.06 | 5.74 | 9.05 | 0.09 | 8.30 | 15.64 | 0.13 | 15.60 |
| FeO | 0.10 | 0.03 | 0.00 | 1.87 | 0.10 | 1.84 | 9.26 | 0.16 | 8.60 | 5.29 | 0.13 | 5.10 |
| TiO$_2$ | 0.03 | 0.03 | 0.00 | 0.07 | 0.03 | 0.00 | 1.15 | 0.16 | 1.01 | 0.57 | 0.09 | 0.37 |
| Total | 100.20 |  | 98.97 | 99.61 |  | 98.98 | 98.62 |  | 91.28 | 99.02 |  | 98.57 |
| N | 6 |  |  | 6 |  |  | 6 |  |  | 6 |  |  |

|  | Moon Mare |  | Start | Earth Cont.Cr. |  | Start | Earth MORB |  | Start | Mars Rego |  | Start | Mars Surf |  | Start |
|---|---|---|---|---|---|---|---|---|---|---|---|---|---|---|---|
| Na$_2$O | 0.04 | 0.03 | 0.00 | 3.10 | 0.11 | 3.07 | 2.76 | 0.06 | 2.78 | 3.68 | 0.14 | 3.56 | 3.07 | 0.10 | 3.03 |
| MgO | 9.00 | 0.12 | 9.40 | 4.25 | 0.14 | 4.66 | 7.91 | 0.09 | 7.95 | 14.76 | 0.26 | 14.33 | 12.03 | 0.14 | 12.19 |
| SiO$_2$ | 44.05 | 0.19 | 44.60 | 61.73 | 0.69 | 60.60 | 50.22 | 0.10 | 50.67 | 59.71 | 0.27 | 59.26 | 51.22 | 0.26 | 50.40 |
| Al$_2$O$_3$ | 10.66 | 0.07 | 10.60 | 15.22 | 0.36 | 15.90 | 15.72 | 0.13 | 15.87 | 13.32 | 0.11 | 13.05 | 11.28 | 0.10 | 11.10 |
| K$_2$O | 0.04 | 0.02 | 0.00 | 1.75 | 0.10 | 1.81 | 0.04 | 0.01 | 0.00 | 0.65 | 0.03 | 0.63 | 0.62 | 0.03 | 0.54 |
| CaO | 10.48 | 0.10 | 10.60 | 5.98 | 0.18 | 6.41 | 11.30 | 0.09 | 11.45 | 6.93 | 0.10 | 6.74 | 5.63 | 0.09 | 5.73 |
| FeO | 18.80 | 0.15 | 19.10 | 6.36 | 0.20 | 6.71 | 9.34 | 0.10 | 9.37 | 0.08 | 0.03 | 0.00 | 13.73 | 0.23 | 14.94 |
| TiO$_2$ | 4.36 | 0.15 | 4.70 | 0.87 | 0.05 | 0.72 | 1.42 | 0.06 | 1.39 | 0.94 | 0.10 | 0.82 | 0.72 | 0.10 | 0.70 |
| Total | 97.43 |  | 99.00 | 99.27 |  | 99.88 | 98.70 |  | 99.49 | 100.06 |  | 98.39 | 98.29 |  | 98.63 |
| N | 6 |  |  | 6 |  |  | 6 |  |  | 6 |  |  | 6 |  |  |

(a)

|  | Merc BS (Vol) | | Merc BS (Mix) | | Merc BS (Chon1) | | Merc BS (Chon2) | | Merc BS (Ref) | Moon BS | | Earth BS | | Mars BS | |
|---|---|---|---|---|---|---|---|---|---|---|---|---|---|---|---|
| Na$_2$O | 0.02 | 0.00 | 0.01 | 0.02 | 0.00 | 0.00 | 0.01 | 0.01 | 0.01 | 0.00 | 0.00 | 0.01 | 0.02 | 0.00 | 0.00 |
| MgO | 55.24 | 0.09 | 56.46 | 0.31 | 56.76 | 0.13 | 57.24 | 0.30 | 55.34 | 55.05 | 0.07 | 56.08 | 0.26 | 54.15 | 0.15 |
| SiO$_2$ | 42.37 | 0.29 | 42.55 | 0.18 | 42.75 | 0.22 | 42.78 | 0.17 | 39.51 | 42.36 | 0.02 | 42.19 | 0.45 | 42.03 | 0.17 |
| Al$_2$O$_3$ | 0.04 | 0.02 | 0.13 | 0.03 | 0.04 | 0.02 | 0.05 | 0.01 | 5.81 | 0.05 | 0.01 | 0.05 | 0.03 | 0.04 | 0.01 |
| K$_2$O | 0.01 | 0.01 | 0.01 | 0.01 | 0.00 | 0.00 | 0.00 | 0.00 | 0.00 | 0.01 | 0.02 | 0.00 | 0.00 | 0.00 | 0.00 |
| CaO | 0.10 | 0.03 | 1.07 | 0.10 | 0.17 | 0.03 | 0.22 | 0.01 | 0.00 | 0.14 | 0.01 | 0.15 | 0.00 | 0.07 | 0.01 |
| FeO | 3.32 | 0.23 | 0.67 | 0.04 | 1.34 | 0.05 | 0.72 | 0.03 | 0.00 | 3.55 | 0.02 | 1.83 | 0.29 | 4.73 | 0.12 |
| TiO$_2$ | 0.02 | 0.02 | 0.06 | 0.03 | 0.02 | 0.02 | 0.01 | 0.02 | 0.02 | 0.01 | 0.01 | 0.03 | 0.01 | 0.02 | 0.03 |
| Total | 101.11 | | 100.95 | | 101.07 | | 101.03 | | 100.69 | 101.15 | | 100.33 | | 101.05 | |
| N | 3 | | 3 | | 3 | | 3 | | 1 | 2 | | 2 | | 3 | |

**(b)**

|  | Moon BS Fass.. | | Merc. BS (Vap) Monti. | | Merc. BS (Ref) Spinel | | Merc BS (Vol) Mferri. | Merc. BS (Chon1) Mferri. Spinel |
|---|---|---|---|---|---|---|---|---|
| Na$_2$O | 0.02 | 0.02 | 0.00 | | 0.01 | 0.01 | 0.00 | 0.01 |
| MgO | 15.92 | 1.24 | 31.53 | | 27.44 | 0.07 | 19.18 | 21.76 |
| SiO$_2$ | 47.25 | 0.27 | 39.30 | | 0.21 | 0.02 | 0.18 | 0.74 |
| Al$_2$O$_3$ | 7.46 | 0.14 | 0.15 | | 71.19 | 0.19 | 2.21 | 13.70 |
| K$_2$O | 0.04 | 0.01 | 0.01 | | 0.00 | 0.01 | 0.00 | 0.02 |
| CaO | 6.97 | 0.27 | 29.67 | | 0.01 | 0.02 | 0.00 | 0.00 |
| FeO | 19.55 | 0.67 | 0.45 | | 0.00 | 0.00 | 69.00 | 54.14 |
| TiO$_2$ | 0.37 | 0.05 | 0.24 | | 0.00 | 0.00 | 0.22 | 1.39 |
| Total | 97.59 | | 101.35 | | 98.87 | | 90.79 | 91.77 |
| N | 3 | | 1 | | 4 | | 1 | 1 |

**(c)**

Table 1a,b: EMPA data (in wt%) with standard deviation. (a) Compositions of glass, Start = composition oft he starting mixture (b) average olivin phases, (c) representative compositions of minor crystalline phases in the glasses. Fass.: Fassaite/Pigeonite, Spinel, Mferri. = Magnesioferrite. Earth Cont.Cr. = Earth Continental Crust, Rego = Regolith., Surf. = Surface.

| Planet Mercury | Grain Size μm | Band Pos. μm | μm | μm | μm | μm | μm | μm | μm | μm | μm | μm | μm | μm | μm | μm | μm | μm | μm | μm | μm | μm | μm | μm |
|---|---|---|---|---|---|---|---|---|---|---|---|---|---|---|---|---|---|---|---|---|---|---|---|---|
| BS (Vol) | 0-25 | | | | | | 4.83 | 5.04 | 5.28 | 5.52 | | | | | 8.33 | | 9.87 | 10.13 | | 11.24 | | 12.41 | | | 16.28 |
| BS (Vol) | 25-63 | | | | | | 4.83 | 5.04 | 5.28 | 5.52 | | | | | 8.16 | | 9.82 | 10.13 | 10.63 | | 11.92 | | | | 16.27 |
| BS (Vol)l | 63-125 | | | | | | 4.83 | | | | | | | | 8.09 | | 9.77 | 10.13 | 10.63 | | 11.92 | | | | 16.24 |
| BS (Vol) | 125-250 | | | | | | | | | | | | | | 8.06 | | 9.82 | 10.13 | 10.58 | | 11.92 | | | | 16.27 |
| BS (Mix) | 0-25 | 2.94 | 3.38 | 3.42 | 3.50 | | 4.00 | 5.01 | 5.26 | 5.50 | 5.75 | | 6.13 | 6.31 | 8.83 | | | 10.12 | 10.62 | 11.14 | | 12.54 | | | 16.18 |
| BS (Mix) | 25-63 | 2.90 | 3.38 | 3.42 | 3.50 | | 4.00 | 5.01 | 5.26 | 5.50 | 5.75 | 5.91 | 6.13 | 6.31 | 8.77 | | | 10.13 | 10.62 | | 11.93 | | | | 16.10 |
| BS (Mix) | 63-125 | 2.87 | 3.38 | 3.42 | 3.50 | | 4.00 | 5.01 | 5.26 | 5.50 | 5.75 | 5.91 | 6.15 | 6.31 | 8.69 | | | 10.13 | 10.57 | | 11.93 | | | | 16.10 |
| BS (Mix) | 125-250 | 2.84 | 3.38 | 3.42 | 3.50 | | 4.00 | 5.01 | 5.26 | 5.50 | 5.75 | | 6.15 | 6.31 | 8.66 | | | 10.13 | 10.57 | | 11.93 | | | | 16.06 |
| BS (Chon1) | 0-25 | 2.93 | 3.38 | 3.41 | 3.50 | | 4.00 | 4.98 | | 5.49 | 5.75 | | 6.13 | | 8.34 | | 9.85 | 10.17 | 10.72 | 11.13 | | 12.35 | | | 16.14 |
| BS (Chon1) | 25-63 | 2.90 | 3.38 | 3.41 | 3.50 | | 4.01 | 4.98 | | 5.49 | 5.75 | 5.90 | 6.13 | | 8.23 | | | 10.17 | 10.57 | | 11.90 | | | | 16.18 |
| BS (Chon1) | 63-125 | 2.84 | 3.38 | 3.41 | 3.50 | | 4.01 | 4.98 | | 5.49 | 5.75 | | 6.13 | | 8.19 | | | 10.17 | 10.57 | | 11.90 | | | | 16.15 |
| BS (Chon1) | 125-250 | 2.82 | 3.38 | 3.41 | 3.50 | | 4.01 | 4.98 | | 5.49 | 5.75 | 5.90 | 6.13 | | 8.17 | | | 10.17 | 10.57 | | 11.90 | | | | 16.12 |
| BS (Chon2) | 0-25 | 2.95 | 3.38 | 3.42 | 3.50 | | | 4.96 | 5.26 | 5.52 | 5.76 | | 6.11 | 6.46 | 8.43 | 9.32 | | 10.14 | 10.48 | | | 12.39 | | 14.01 | |
| BS (Chon2) | 25-63 | 2.93 | 3.38 | 3.42 | 3.50 | | | 4.96 | 5.26 | 5.52 | 5.76 | | 6.11 | 6.50 | 8.31 | 9.32 | | 10.14 | 10.45 | | | | 13.60 | | |
| BS (Chon2) | 63-125 | 2.91 | 3.38 | 3.42 | 3.50 | | | 4.96 | 5.26 | 5.52 | 5.76 | | 6.11 | 6.50 | 8.23 | 9.32 | | 10.14 | 10.45 | | 10.79 | | 13.60 | | |
| BS (Chon2) | 125-250 | 2.91 | 3.38 | 3.42 | 3.50 | | | 4.96 | 5.26 | 5.52 | 5.76 | | 6.11 | 6.50 | 8.20 | 9.32 | | 10.14 | 10.45 | | 10.79 | | 13.60 | | |
| BS (Vap) | 0-25 | 2.97 | 3.38 | 3.42 | 3.50 | | | 4.99 | 5.28 | | 5.65 | 5.88 | 6.02 | | | | | 10.27 | 10.67 | | | 12.63 | 13.91 | 14.30 | 16.39 |
| BS (Vap) | 25-63 | 2.94 | 3.38 | 3.42 | 3.50 | | 4.03 | 4.99 | 5.28 | | 5.65 | 5.88 | 6.02 | | 8.97 | 9.48 | | 10.29 | 10.67 | | | | | 14.36 | 16.30 |
| BS (Vap) | 63-125 | 2.94 | 3.38 | 3.42 | 3.50 | | 4.03 | 4.99 | 5.28 | | 5.65 | 5.88 | 6.02 | | 8.96 | 9.50 | | 10.29 | 10.68 | | | | | 14.36 | 16.27 |
| BS (Vap) | 125-250 | 2.94 | 3.38 | 3.42 | 3.50 | 3.83 | 4.03 | 4.99 | 5.28 | | 5.65 | 5.88 | 6.02 | | 8.59 | 8.94 | 9.52 | 10.29 | 10.71 | | | | | 14.41 | 16.32 |
| BS (Ref)f | 0-25 | 2.90 | 3.38 | 3.42 | 3.50 | | 4.01 | | | 5.50 | 5.74 | | 6.08 | 6.31 | 8.63 | 9.33 | | 10.12 | 10.59 | 11.35 | | 12.13 | | 14.18 | |
| BS (Ref)f | 25-63 | 2.88 | 3.38 | 3.42 | 3.50 | | 4.01 | | | 5.50 | 5.74 | | 6.08 | 6.39 | 8.61 | 9.33 | | 10.12 | 10.51 | | 11.92 | | | 14.18 | |
| BS (Ref)f | 63-125 | 2.89 | 3.38 | 3.42 | 3.50 | | 4.01 | | | 5.50 | 5.74 | | 6.08 | 6.39 | 8.59 | 9.33 | | 10.12 | 10.51 | | 11.92 | | | 14.22 | |
| BS (Ref) | 125-250 | 2.89 | 3.38 | 3.42 | 3.50 | | 4.01 | | | 5.50 | 5.74 | | 6.08 | 6.39 | 8.63 | 9.33 | | 10.12 | 10.51 | | 11.92 | | 13.68 | 14.10 | |

| Planet | Grain Size | Band Pos. | | | | | | | | | | | | | | | | | | | | | | | |
|---|---|---|---|---|---|---|---|---|---|---|---|---|---|---|---|---|---|---|---|---|---|---|---|---|---|
| Mercury | µm | µm | µm | µm | µm | µm | µm | µm | µm | µm | µm | µm | µm | µm | µm | µm | µm | µm | µm | µm | µm | µm | µm | µm | µm |
| Regolith | 0-25 | 2.94 | 3.38 | 3.42 | 3.50 | | | | | | | | | | | 8.07 | 9.81 | | | 11.95 | | | | |
| Regolith | 25-63 | 2.90 | 3.38 | 3.42 | 3.50 | | | | | | | | | | | 8.04 | 9.84 | | | | | | | |
| Regolith | 63-125 | 2.85 | 3.38 | 3.42 | 3.50 | | | | | | | | | | | 8.04 | 9.81 | | | | | | | |
| Regolith | 125-250 | 2.83 | 3.38 | 3.42 | 3.50 | | | | | | | | | | | 8.04 | 9.81 | | | | | | | |
| | | | | | | | | | | | | | | | | | | | | | | | | | |
| Surface | 0-25 | 2.93 | 3.38 | 3.42 | 3.50 | | | | | | | | | | | 8.09 | 9.77 | | | 11.98 | | | | |
| Surface | 25-63 | 2.86 | 3.38 | 3.42 | 3.50 | | | | | | | | | | | 8.08 | 9.83 | | | | | | | |
| Surface | 63-125 | 2.83 | 3.38 | 3.42 | 3.50 | | | | | | | | | | | 8.06 | 9.87 | | | | | | | |
| Surface | 125-250 | 2.82 | 3.38 | 3.42 | | | | | | | | | | | | 8.06 | 9.87 | | | | | | | |

Tab.2a

| Planet | Grain Size | Band Pos. |
|---|---|---|

| Venus | µm | µm | µm | µm | µm | µm | µm | µm | µm | µm | µm | µm | µm | µm | µm | µm | µm | µm | µm | µm | µm | µm | µm | µm | µm | µm |
|---|---|---|---|---|---|---|---|---|---|---|---|---|---|---|---|---|---|---|---|---|---|---|---|---|---|---|
| Surface | 0-25 | 2.91 | 3.38 | 3.42 | 3.50 | | | | | | | | | | | | 8.10 | | 10.57 | | | | | | | |
| Surface | 25-63 | 2.85 | 3.38 | 3.42 | 3.50 | | | | | | | | | | | | 8.10 | | 10.51 | | | | | | | |
| Surface | 63-125 | 2.82 | 3.38 | 3.42 | | | | | | | | | | | | | 8.09 | | 10.56 | | | | | | | |
| Surface | 125-250 | 2.82 | | 3.42 | | | | | | | | | | | | | 8.09 | | 10.53 | | | | | | | |
| **Moon** | | | | | | | | | | | | | | | | | | | | | | | | | | |
| BS | 0-25 | | 3.37 | 3.41 | | | 4.94 | | 5.45 | 5.62 | 5.99 | | | | | | 8.54 | | 10.12 | | | 11.25 | | 12.30 | | 16.34 |
| BS | 25-63 | | | | | | 4.94 | | 5.45 | 5.62 | 5.99 | | | | | | 8.28 | | 10.13 | | 10.75 | | 11.92 | | | 16.27 |
| BS | 63-125 | | | | | | 4.94 | | 5.45 | 5.62 | 5.99 | | | | | | 8.22 | | 10.13 | | 10.75 | | 11.92 | | | 16.24 |
| BS | 125-250 | | | | | | 4.94 | | 5.45 | 5.62 | 5.99 | | | | | | 8.20 | | 10.13 | | 10.75 | | 11.92 | | | 16.31 |
| Highlands | 0-25 | 2.94 | 3.38 | 3.42 | 3.50 | | | | | | | | | | | | 8.19 | | 10.57 | | | | | | | |
| Highlands | 25-63 | 2.87 | 3.38 | 3.42 | 3.50 | | | | | | | | | | | | 8.18 | | 10.50 | | | | | | 14.92 | |
| Highlands | 63-125 | 2.85 | 3.38 | 3.42 | 3.50 | | | | | | | | | | | | 8.18 | | 10.47 | | | | | | 14.73 | |
| Highlands | 125-250 | 2.84 | 3.38 | 3.42 | 3.50 | | | | | | | | | | | | 8.18 | | 10.49 | | | | | | 14.64 | |
| Mare | 0-25 | 2.86 | 3.37 | 3.42 | 3.50 | | | | | | | | | | | | 8.26 | | 10.59 | | | | | | | |
| Mare | 25-63 | 2.82 | 3.37 | 3.42 | 3.50 | | | | | | | | | | | | 8.24 | | 10.56 | | | | | | | |
| Mare | 63-125 | 2.80 | | 3.42 | | | | | | | | | | | | | 8.24 | | 10.56 | | | | | | | |
| Mare | 125-250 | 2.80 | | | | | | | | | | | | | | | 8.23 | | 10.56 | | | | | | | |

Tab. 2b

| Planet | Grain Size | Band | Pos. |
|---|---|---|---|

**Earth**

| Sample | Size | μm | μm | μm | μm | μm | μm | μm | μm | μm | μm | μm | μm | μm | μm | μm | μm |
|---|---|---|---|---|---|---|---|---|---|---|---|---|---|---|---|---|---|
| BS | 0-25 | 2.86 | 3.37 | 3.41 | | | 4.93 | | 5.62 | 5.99 | 8.57 | | 9.73 | 10.12 | | 11.20 | | 12.45 | 16.35 |
| BS | 25-63 | 2.79 | 3.37 | 3.41 | | | 4.93 | | 5.62 | 5.99 | 8.27 | | 9.73 | 10.13 | 10.65 | | 11.93 | | 16.29 |
| BS | 63-125 | 2.78 | | 3.41 | | | 4.93 | | 5.62 | 5.99 | 8.21 | | | 10.13 | 10.62 | | 11.92 | | 16.19 |
| BS | 125-250 | | | | | | 4.93 | | 5.62 | 5.99 | 8.19 | | | 10.13 | 10.61 | | 11.92 | | 16.19 |
| Cont.Crust | 0-25 | 2.91 | 3.38 | 3.42 | 3.50 | 3.98 | | | | | 7.90 | 9.51 | | | | | 11.69 | | |
| Cont.Crust | 25-63 | 2.83 | 3.38 | 3.42 | 3.50 | | | | | | 7.86 | 9.51 | | | | | | | |
| Cont.Crust | 63-125 | 2.82 | 3.38 | 3.42 | | | | | | | 7.88 | 9.50 | | | | | | | |
| Cont.Crust | 125-250 | 2.81 | 3.38 | 3.42 | | | | | | | 7.88 | 9.53 | | | | | | | |
| MORB | 0-25 | 2.93 | 3.37 | 3.41 | 3.50 | | | | | | 8.17 | | | 10.22 | | | 11.80 | | |
| MORB | 25-63 | 2.85 | 3.37 | 3.41 | | | | | | | 8.13 | | | 10.15 | | | | | |
| MORB | 63-125 | 2.83 | | 3.41 | | | | | | | 8.13 | | | 10.15 | | | | | |
| MORB | 125-250 | 2.82 | | | | | | | | | 8.14 | | | 10.15 | | | | | |

**Mars**

| Sample | Size | μm | μm | μm | μm | μm | μm | μm | μm | μm | μm | μm | μm | μm | μm | μm | μm |
|---|---|---|---|---|---|---|---|---|---|---|---|---|---|---|---|---|---|
| BS | 0-25 | | | | | | 4.93 | | 5.62 | 5.99 | 8.25 | | 9.83 | 10.13 | | 11.31 | | 12.41 | 16.39 |
| BS | 25-63 | | | | | | 4.93 | | 5.61 | 5.99 | 8.08 | | 9.74 | 10.14 | 10.71 | | 11.92 | | 16.35 |
| BS | 63-125 | | | | | | | | 5.62 | 5.98 | 8.04 | | 9.76 | 10.14 | 10.68 | | 11.92 | | 16.35 |
| BS | 125-250 | | | | | | | | 5.61 | 6.00 | 8.03 | | 9.74 | 10.14 | 10.71 | | 11.92 | | 16.38 |
| Regolith | 0-25 | 2.93 | 3.38 | 3.41 | 3.50 | | | | | | 7.97 | 9.71 | | | | | 11.86 | | |
| Regolith | 25-63 | 2.89 | 3.38 | 3.41 | 3.50 | | | | | | 7.99 | 9.68 | | | | | | | |
| Regolith | 63-125 | 2.84 | 3.38 | 3.41 | 3.50 | | | | | | 8.01 | 9.74 | | | | | | | |
| Regolith | 125-250 | 2.82 | 3.38 | 3.41 | 3.50 | | | | | | 8.01 | 9.70 | | | | | | | |
| Surface | 0-25 | 2.91 | 3.38 | 3.41 | 3.50 | | | | | | 8.19 | | | 10.01 | | | 11.87 | | |
| Surface | 25-63 | 2.84 | 3.38 | 3.41 | | | | | | | 8.16 | | | 9.96 | | | | | |
| Surface | 63-125 | 2.82 | 3.38 | 3.41 | | | | | | | 8.15 | | | 10.03 | | | | | |
| Surface | 125-250 | 2.82 | | | | | | | | | 8.15 | | | 9.98 | | | | | |

Tab.2c

Table 2a-c: Band positions ( Band Pos.) of spectral features in µm. Grain Size=grain size of sieved fraction analyzed. BS = Bulk Silicate, Cont.Crust = Continental Crust.

| Planet | | |
|---|---|---|
| Venus Surface1 | 8.05 | 10.64 |
| Venus Surface2 | 8.05 | 10.58 |
| Moon Highland1 | 8.19 | 10.58 |
| Moon Highland2 | 8.19 | 10.59 |
| Moon Mare3 | 8.28 | 10.64 |
| Moon Mare4 | 8.26 | 10.64 |

Table 3: Results of micro-FTIR analyses. Band positions in µm.

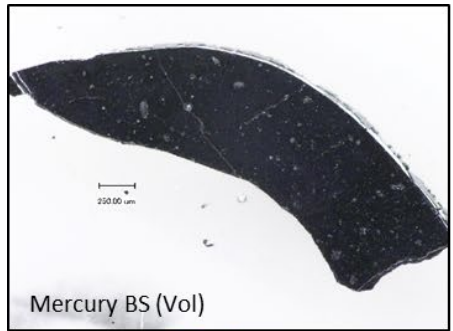
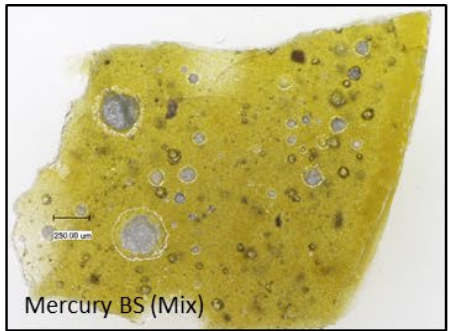
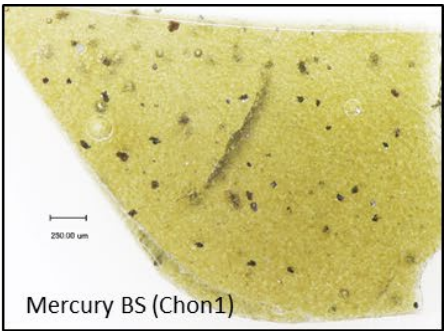
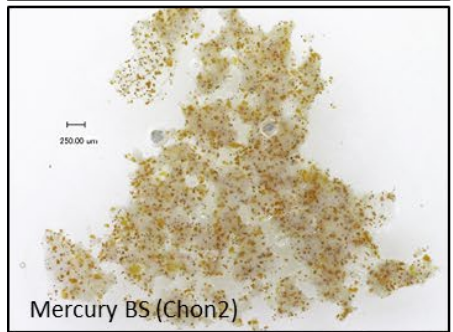
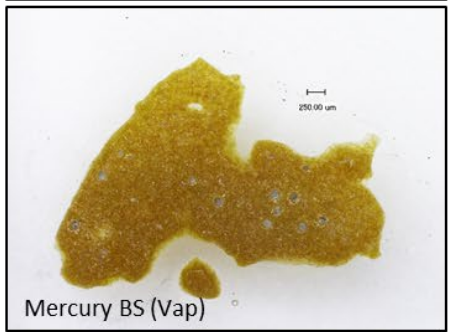
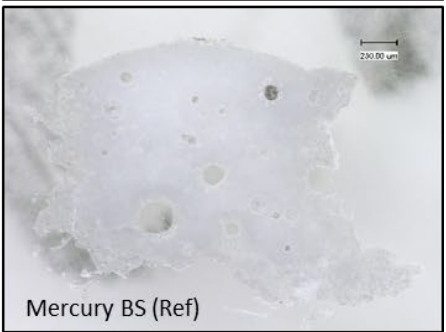
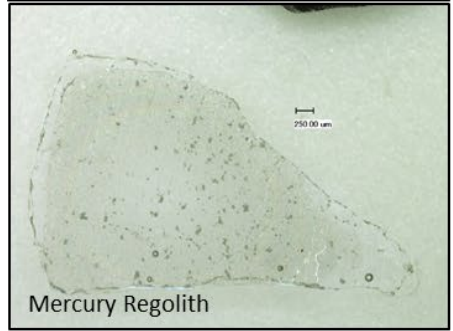
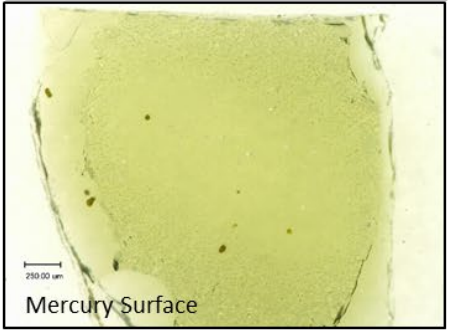
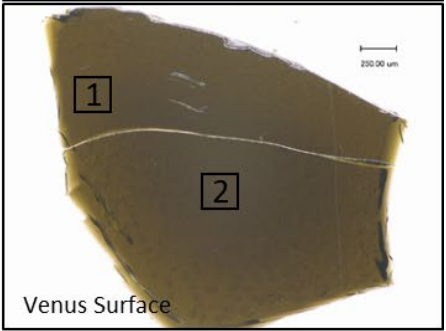
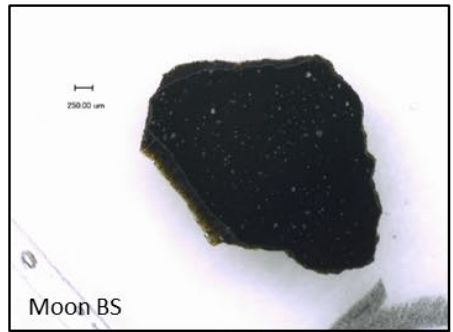
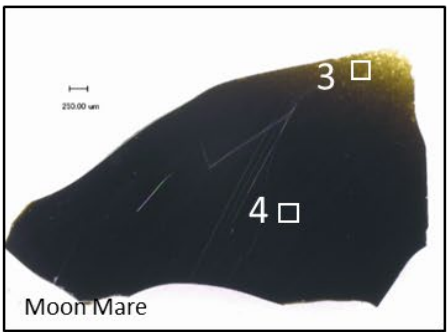
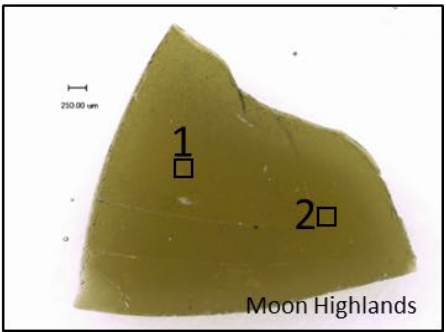
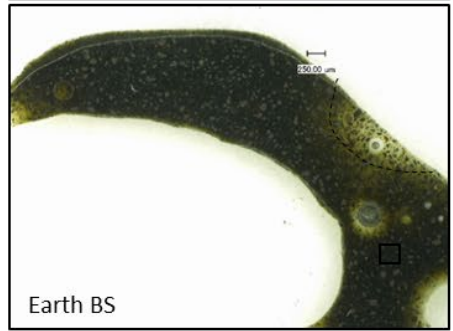
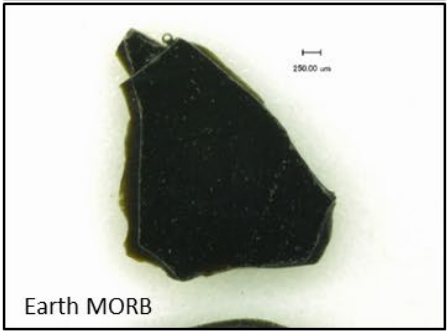
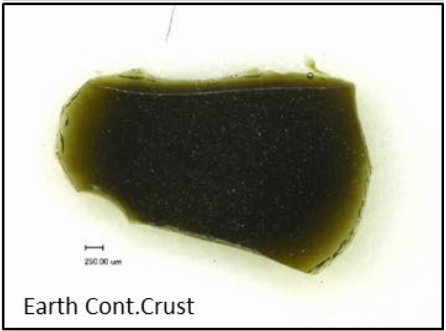
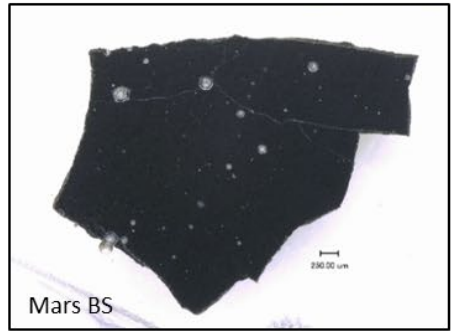
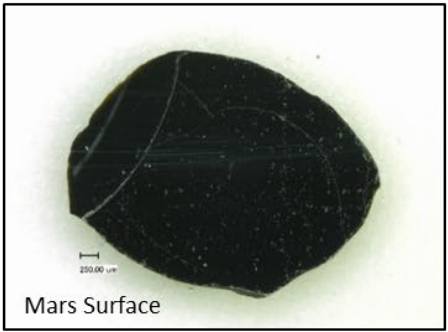
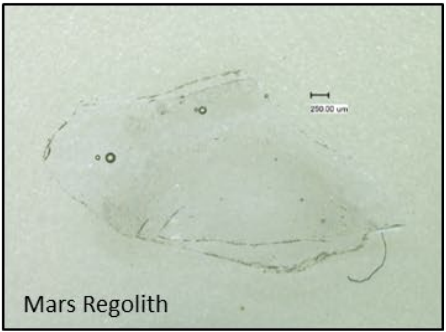

Fig.1a

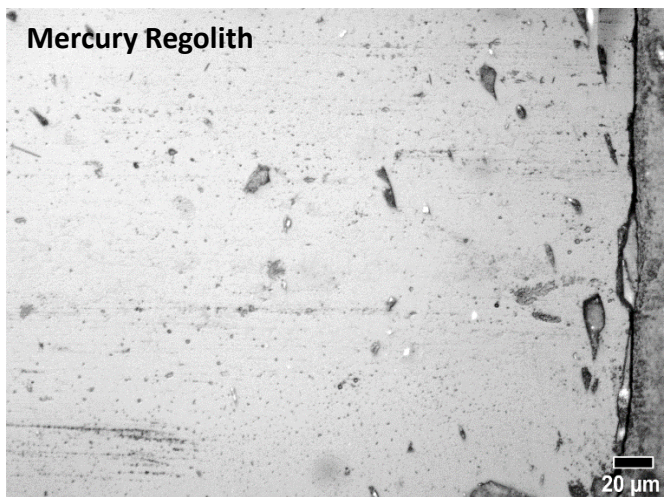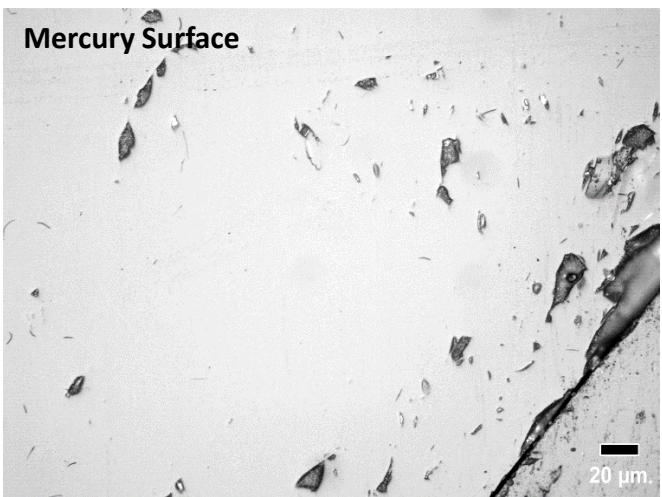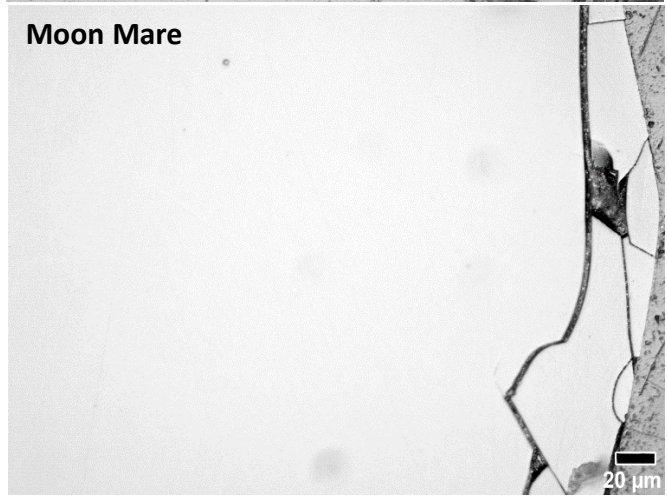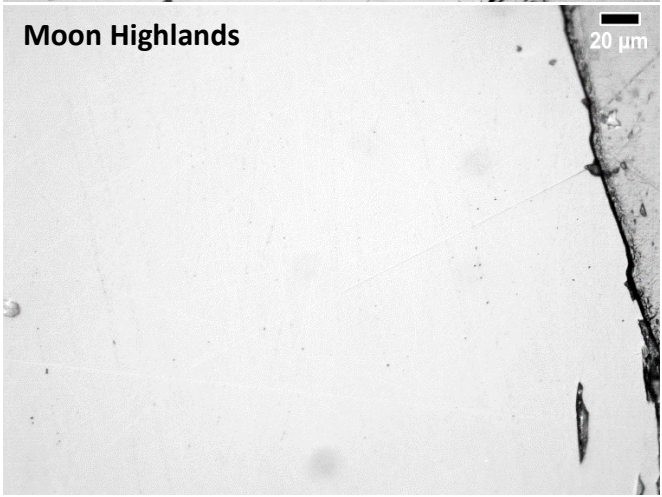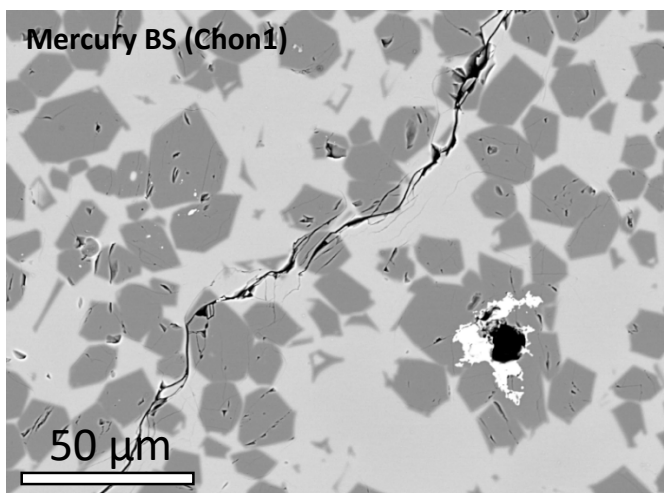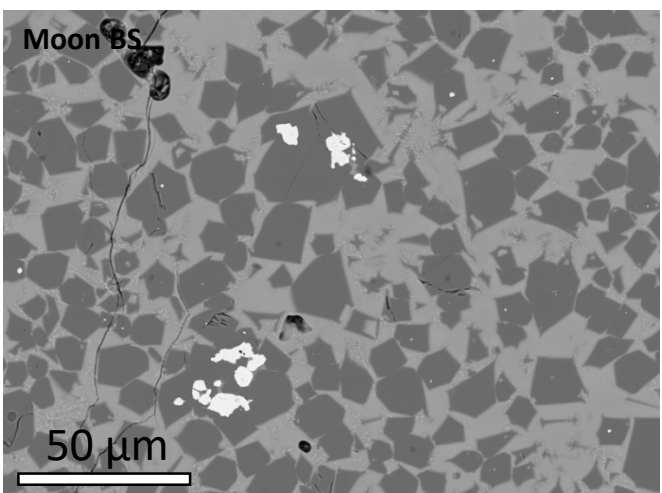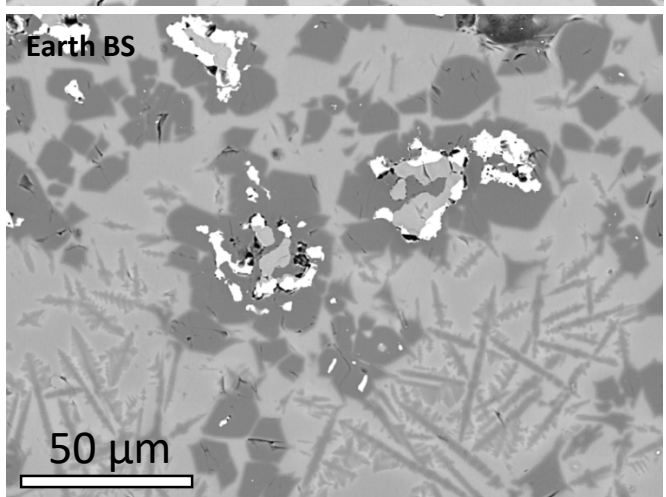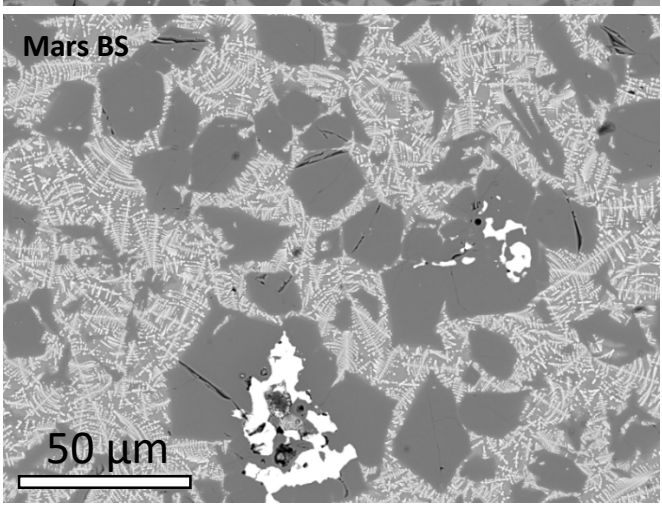

Fig.1b

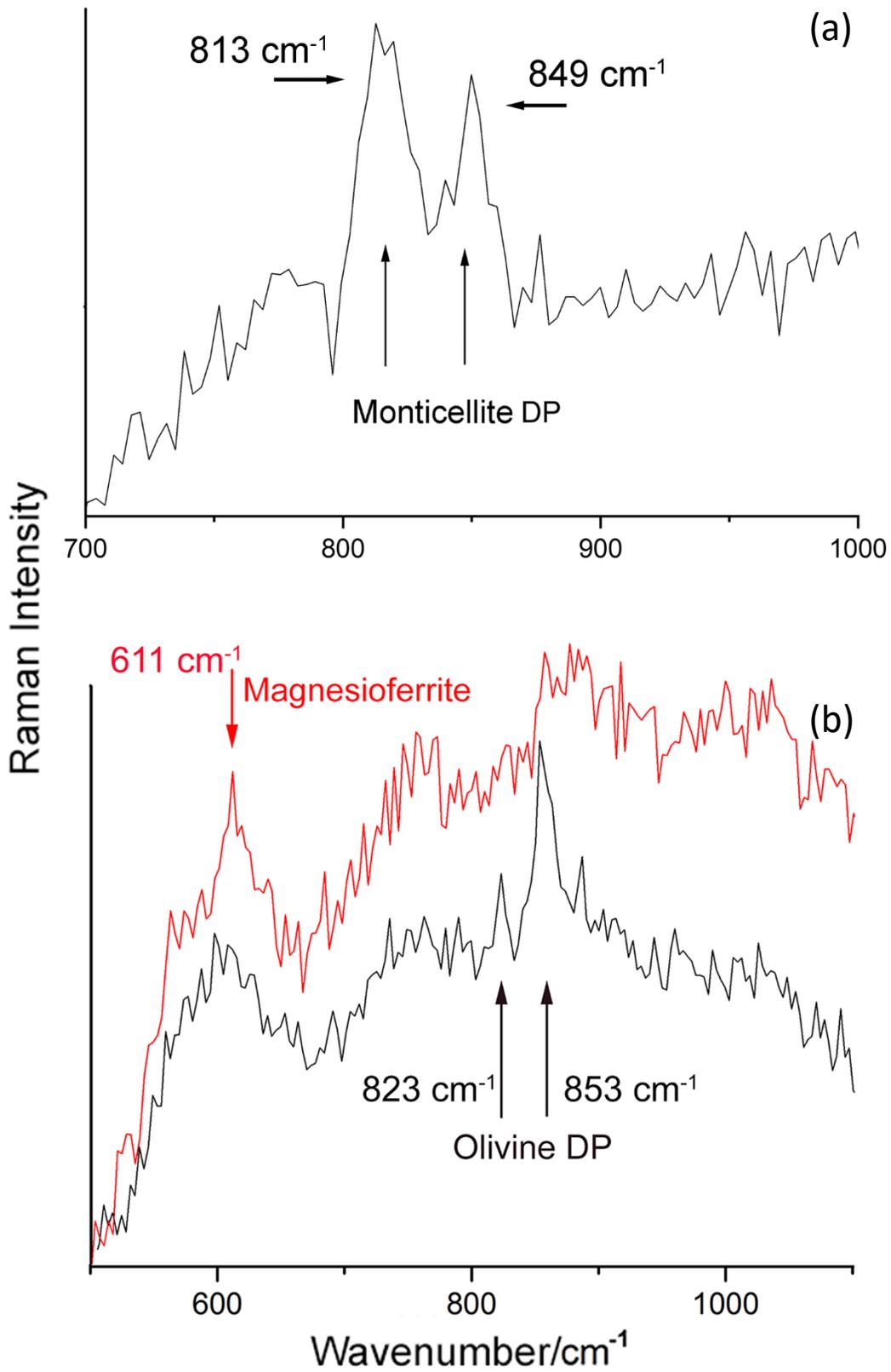

Fig.2

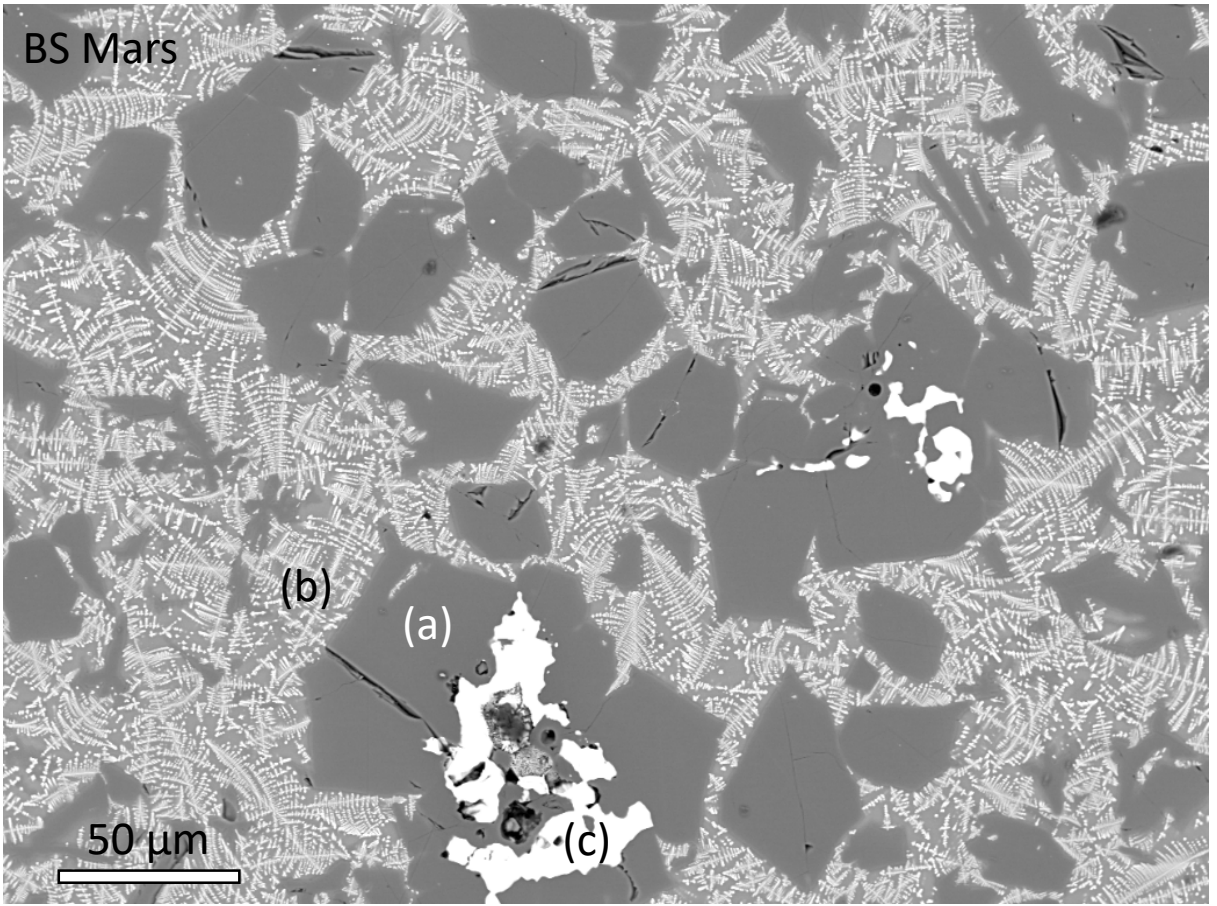

Fig.3

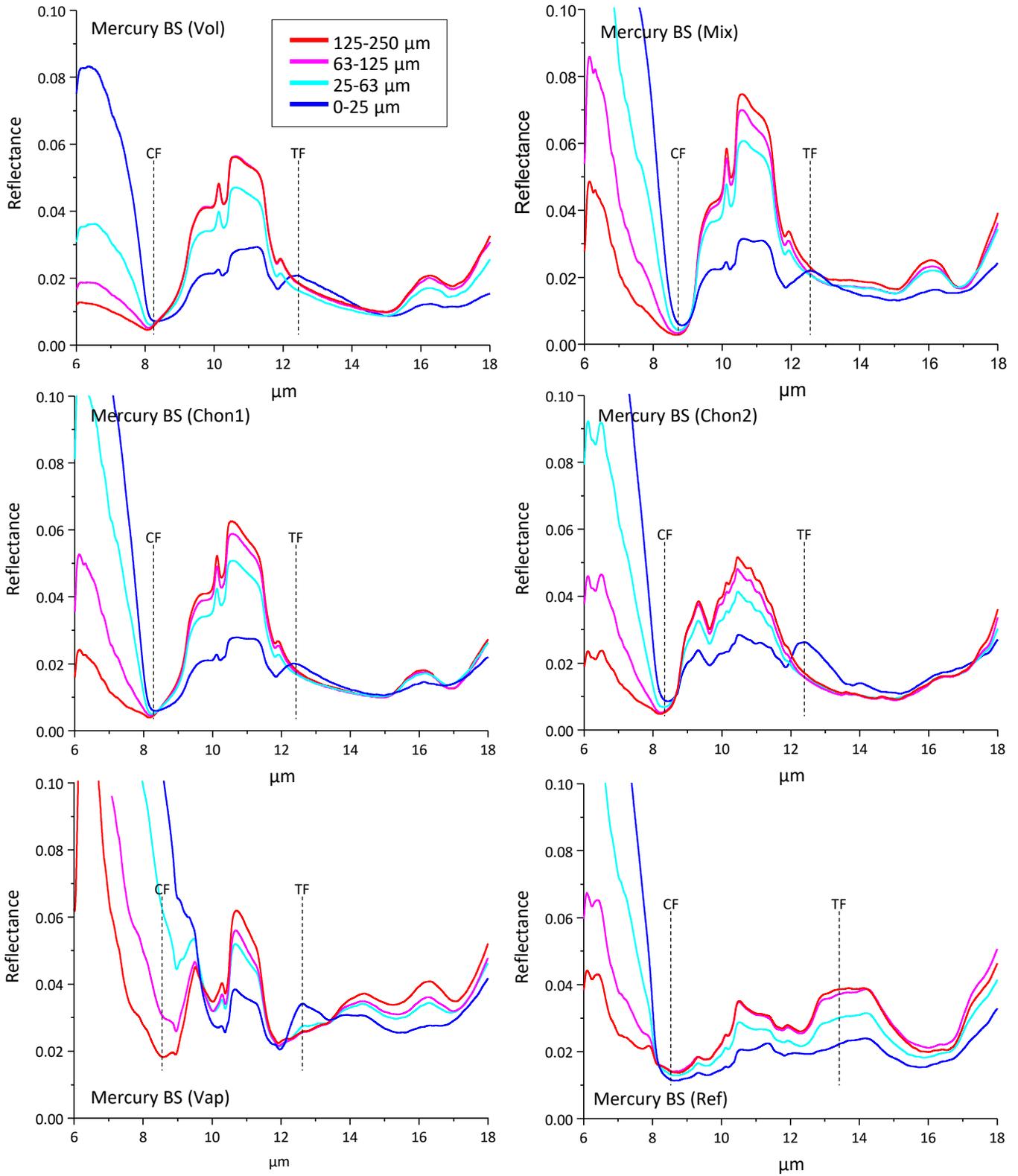

Fig.4a

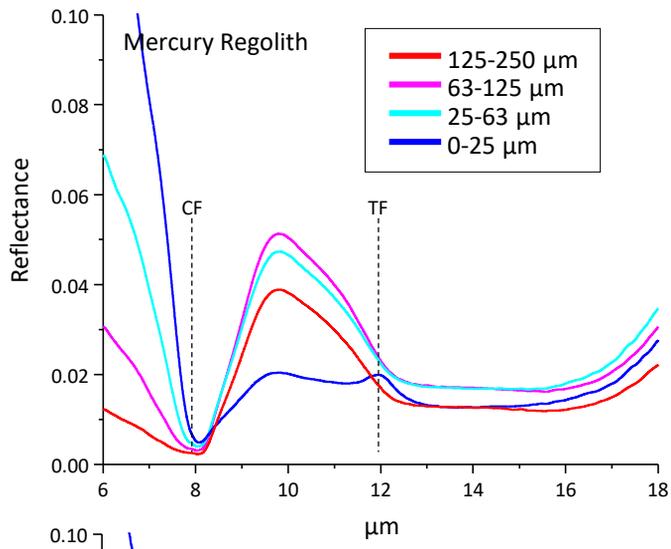
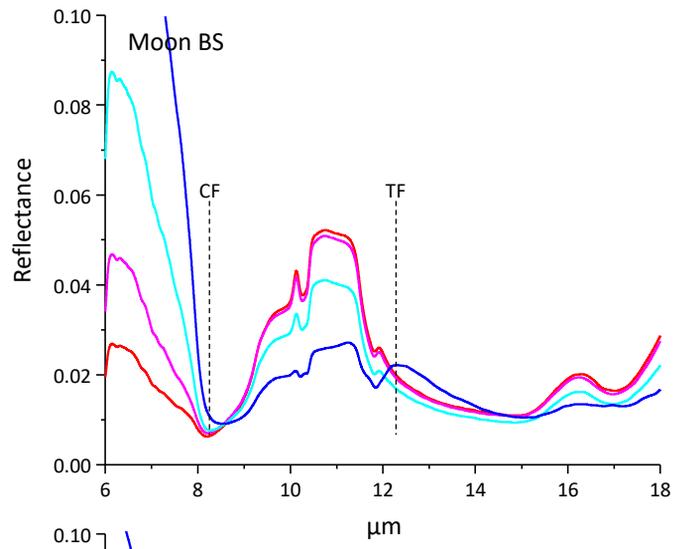
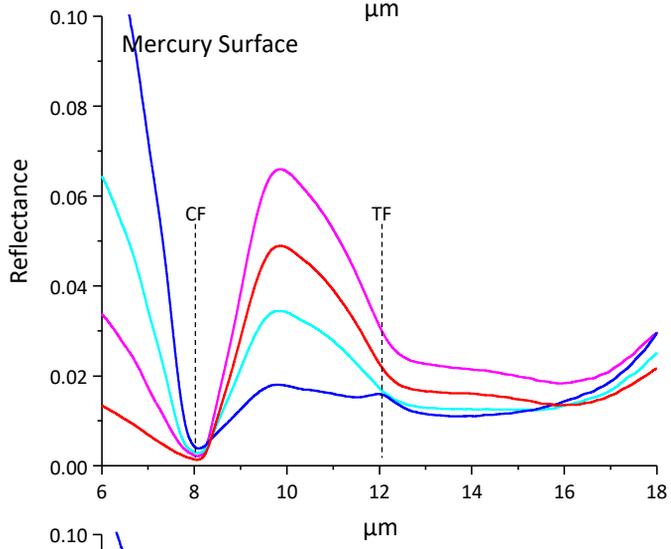
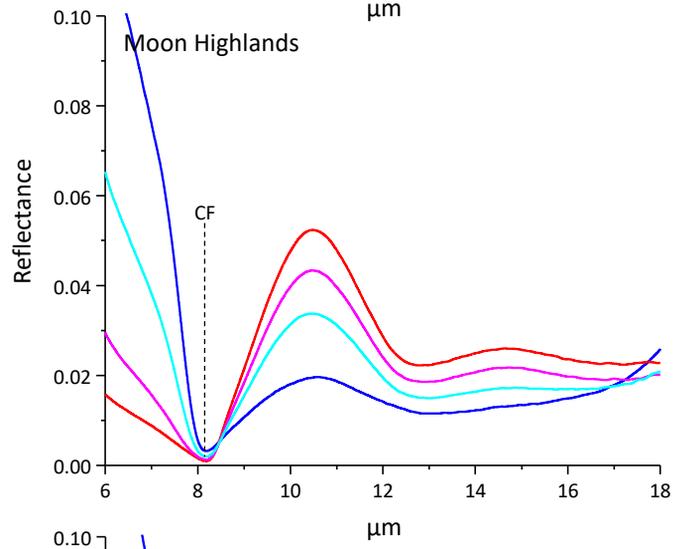
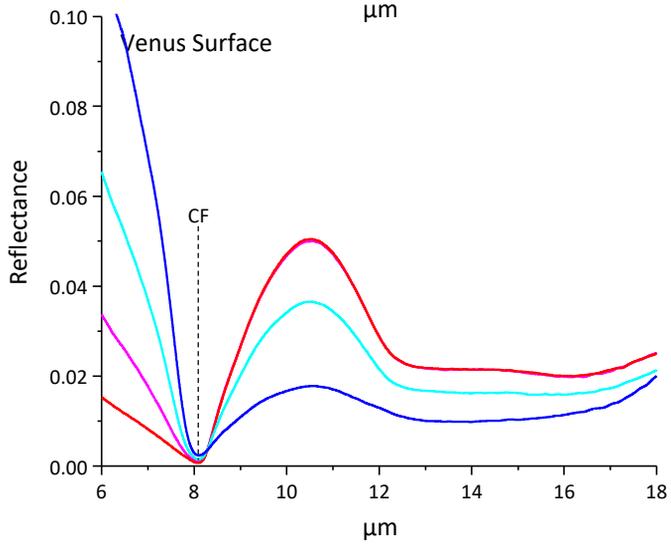
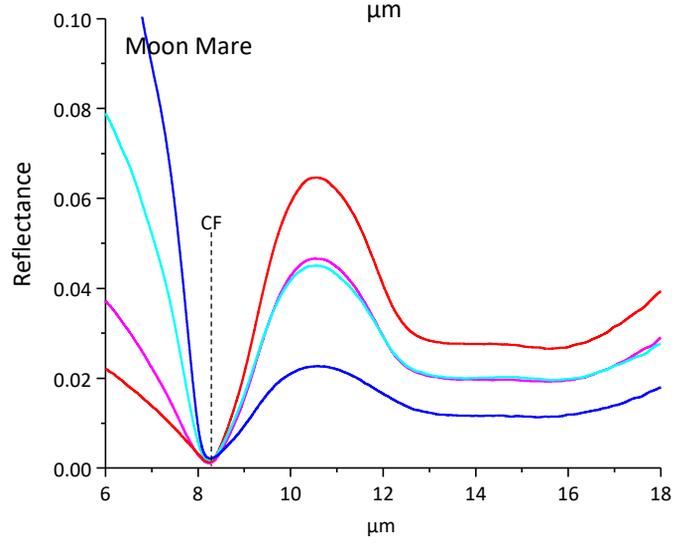

Fig.4b

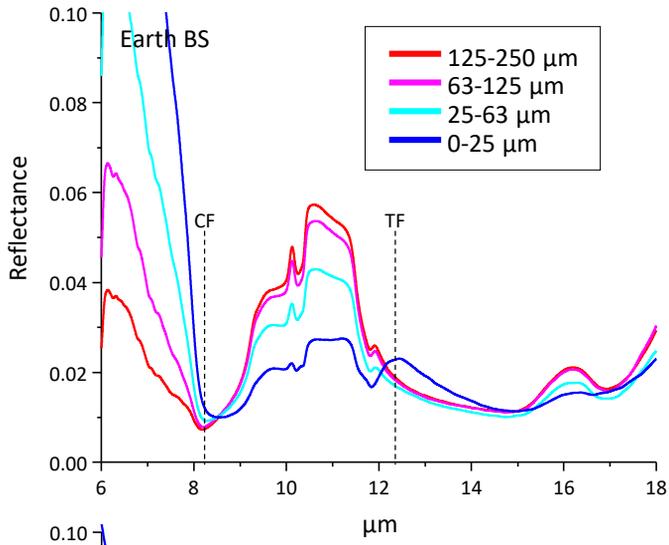
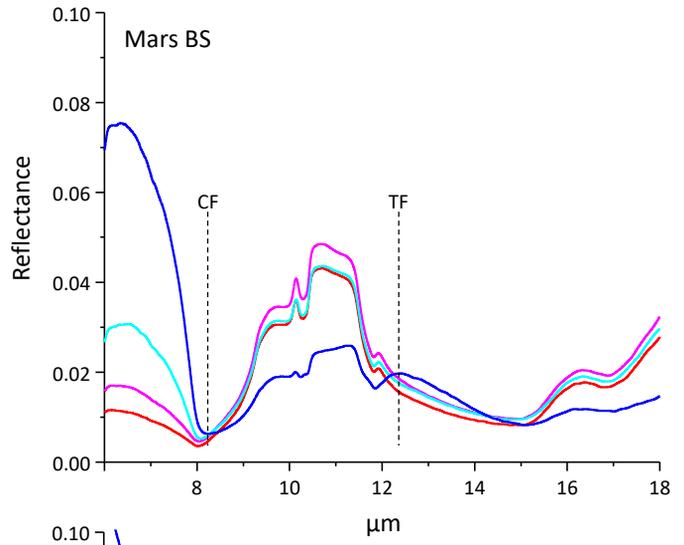
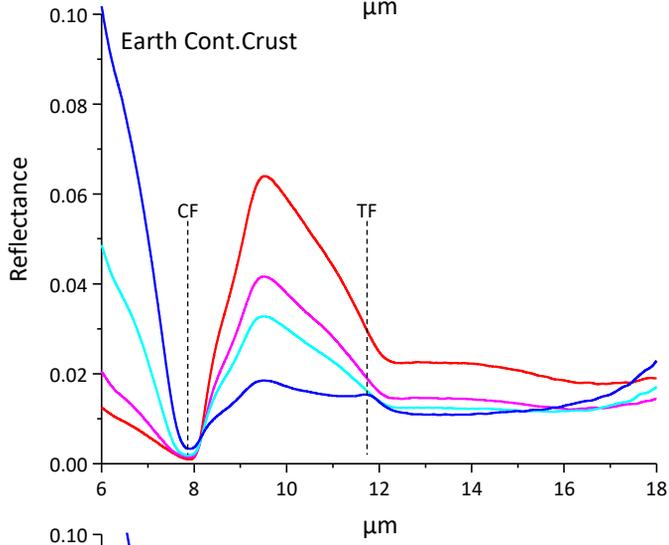
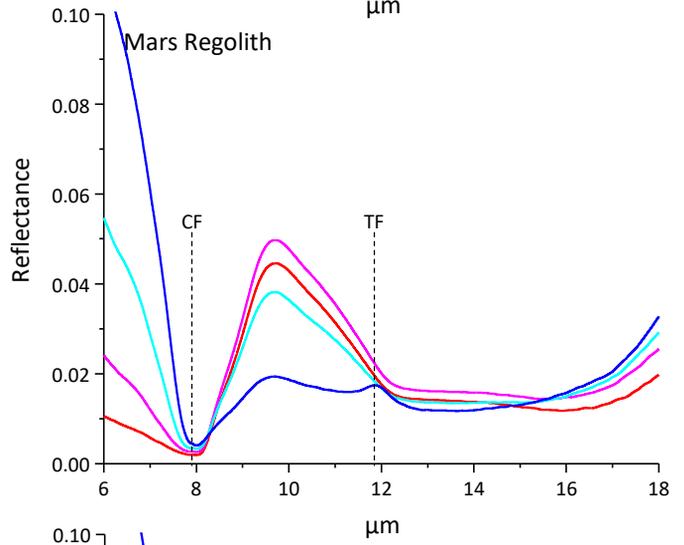
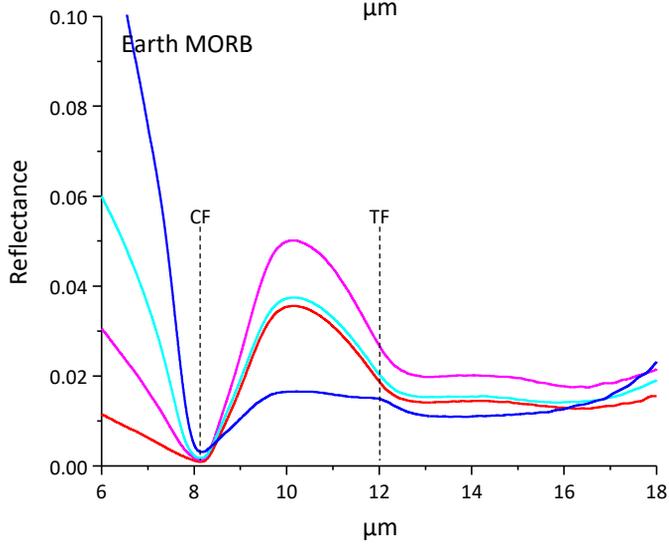
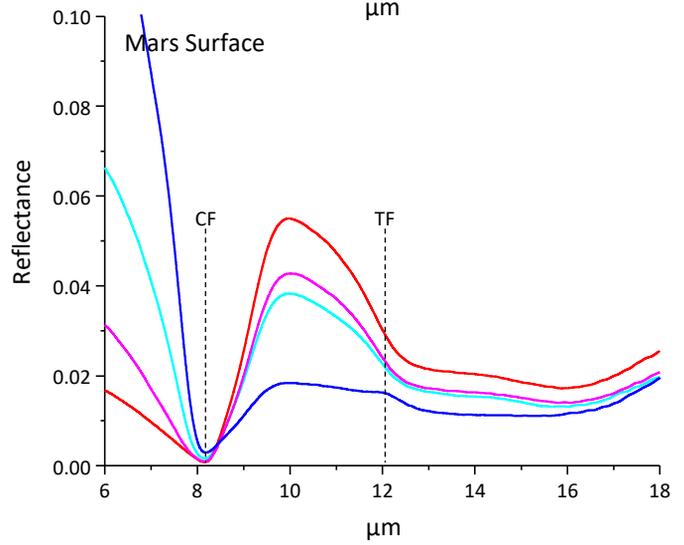

Fig.4c

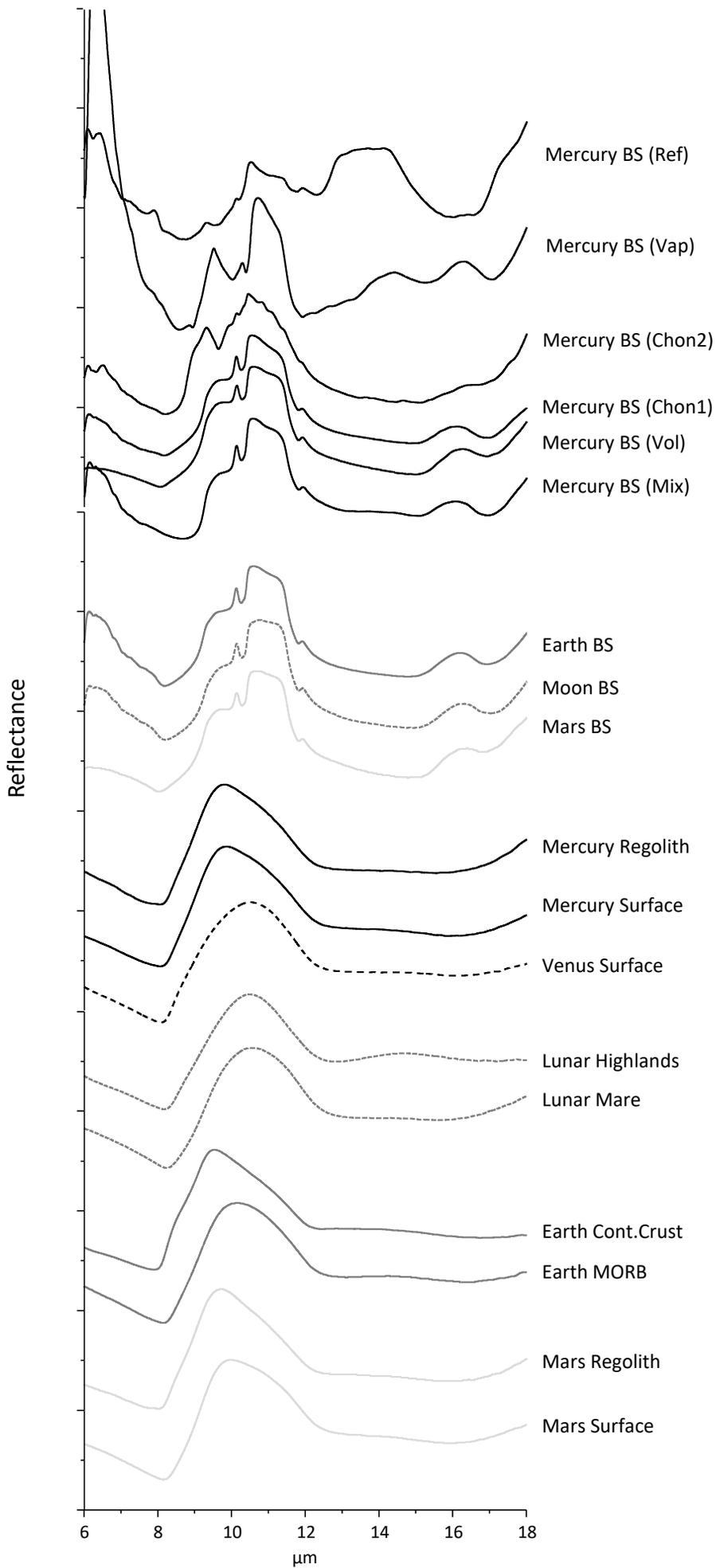

Fig.5a

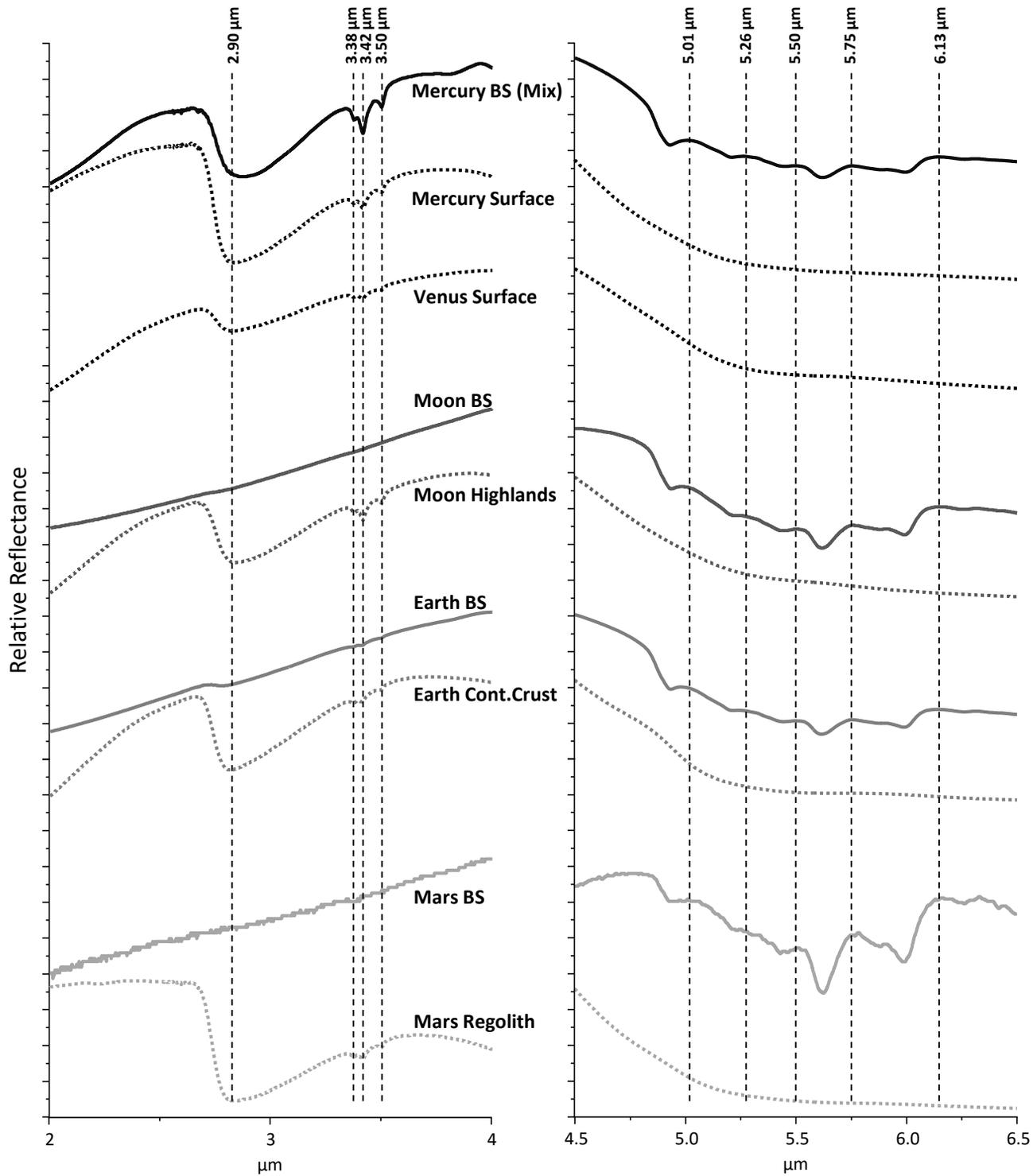

Fig.5b

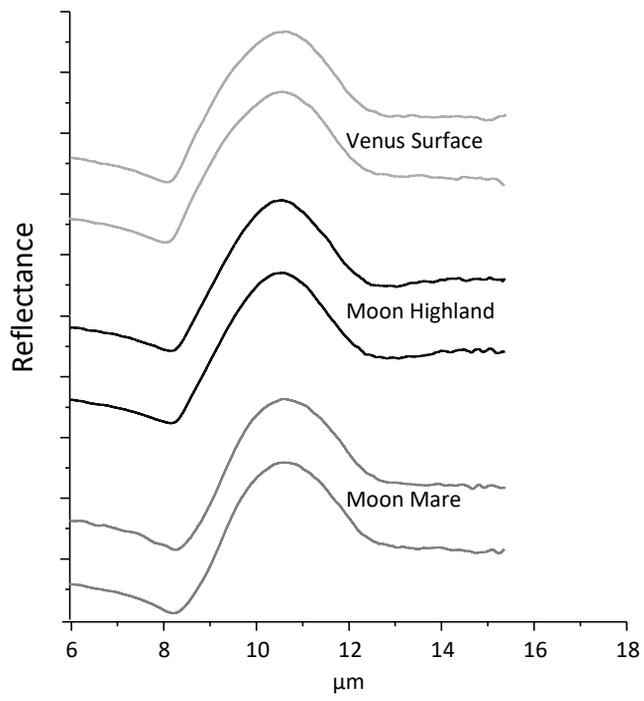

Fig.6

Fig.7

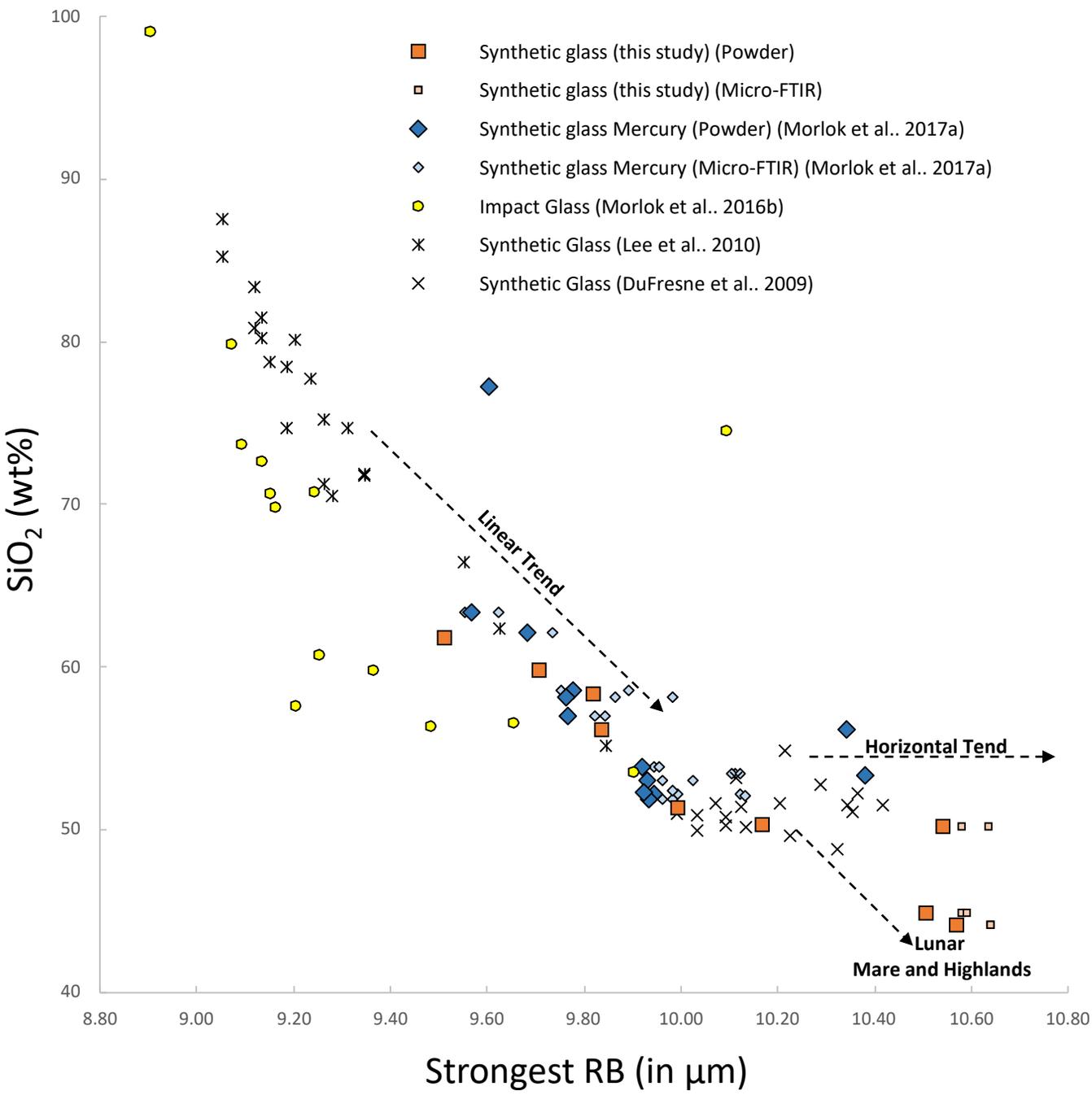

Fig.8

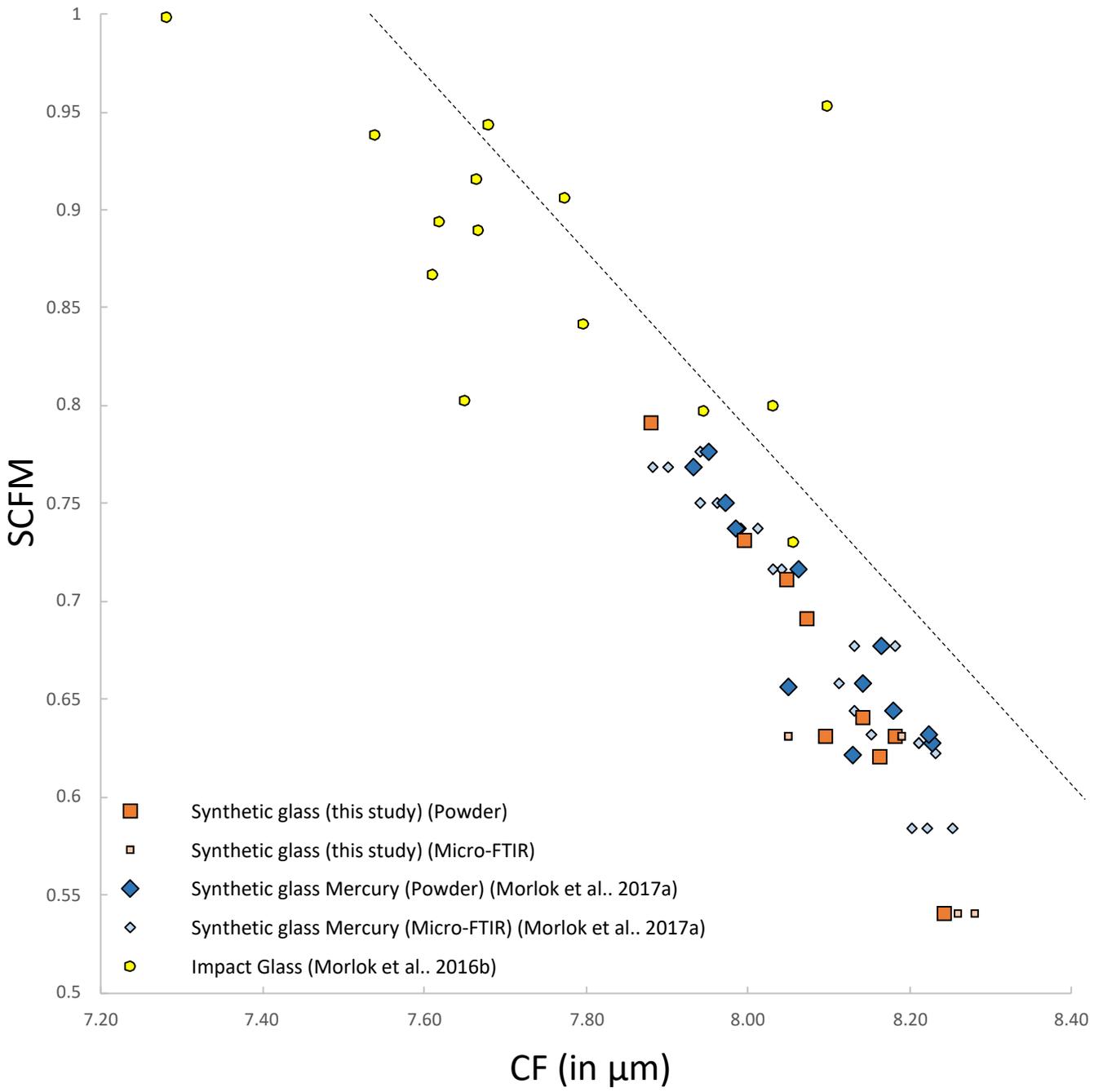

Fig.9

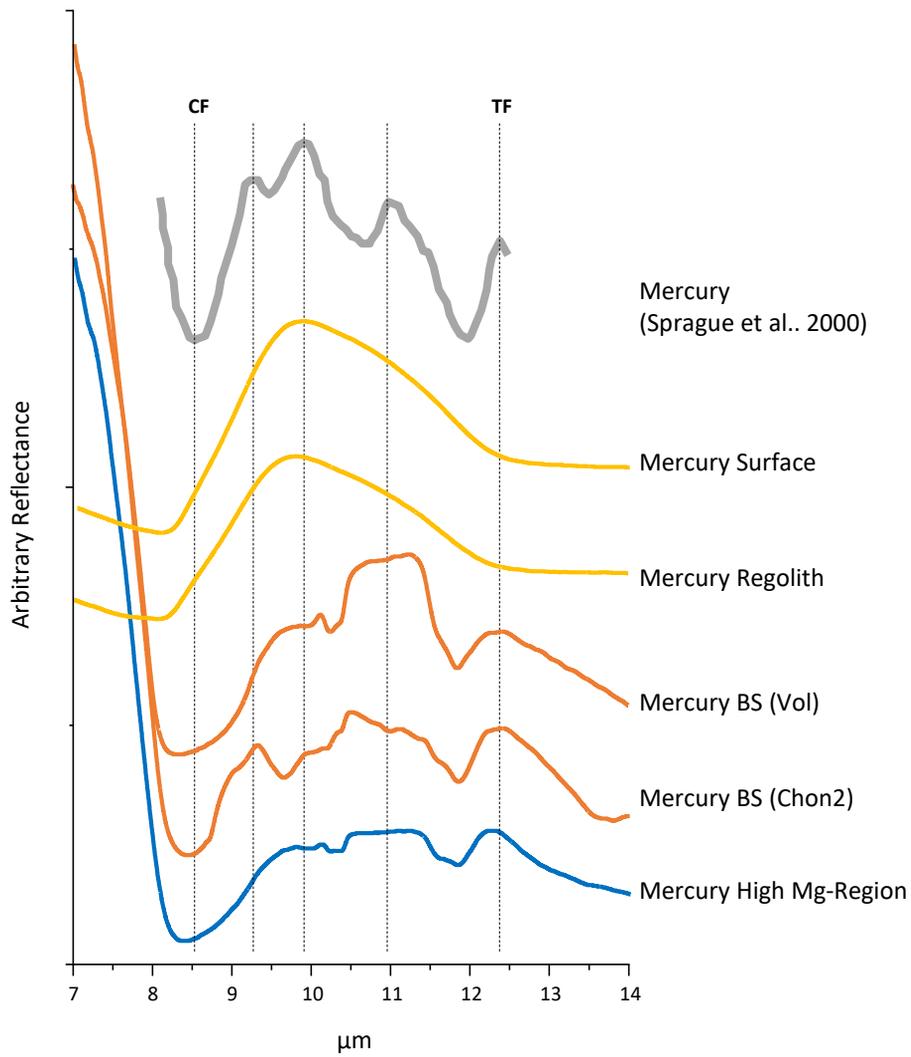

Fig.10